%% file: FileCombine.tex
\documentclass[10pt,journal,compsoc]{IEEEtran}
\usepackage{balance}

\usepackage[utf8]{inputenc}

\title{SE Factual Knowledge in Frozen Giant Code Model: A Study on FQN and its Retrieval}
\author{Qing~Huang,
        Dianshu~Liao,
        Zhenchang~Xing,
        Zhiqiang~Yuan,
        Qinghua~Lu,
        Xiwei~Xu,
        Jiaxing~Lu% <-this % stops a space
\IEEEcompsocitemizethanks{\IEEEcompsocthanksitem Q. Huang, D. Liao, Z. Yuan, J. Lu are with School of Computer Information Engineering, Jiangxi Normal University, China.\protect
\IEEEcompsocthanksitem Q. Huang and D. Liao are co-first authors, Z. Yuan is the corresponding author(yuanzhiq@jxnu.edu.cn)
\IEEEcompsocthanksitem Z. Xing is with Data61 Australia and Australian National University Australia.
\IEEEcompsocthanksitem Q. Lu and X. Xu are with Data61 Australia.}% <-this % stops an unwanted space
}

\date{December 2022}
\usepackage{ulem}
\usepackage{graphicx}
\usepackage{float}
\usepackage{array}
\usepackage{tabularx}
\usepackage{multirow}
\usepackage{amsmath}
\usepackage{amssymb}
\usepackage{threeparttable}
\usepackage{setspace}
\usepackage{color}
\usepackage{colortbl}
\usepackage{xcolor}
\usepackage{soul}
\usepackage{booktabs}
\usepackage{makecell}
\usepackage{algorithm} 
\usepackage{algpseudocode}
\usepackage{hyperref}

\usepackage{verbatim}

\usepackage{picins}

\definecolor{mygray}{gray}{.9}
\definecolor{mypink}{rgb}{.99,.91,.95}
\definecolor{mycyan}{cmyk}{.3,0,0,0}
\definecolor{myyellow}{RGB}{255,230,204}
\definecolor{mybule}{RGB}{218,232,252}
\definecolor{mygreen}{RGB}{213,232,212}
\definecolor{titleColor}{RGB}{102,102,102}

\begin{document}
\IEEEtitleabstractindextext{
\begin{abstract}
Pre-trained giant code models (PCMs) start coming into the developers' daily practices.
Understanding what types of and how much software knowledge is packed into PCMs is the foundation for incorporating PCMs into software engineering (SE) tasks and fully releasing their potential.
In this work, we conduct the first systematic study on the SE factual knowledge in the state-of-the-art PCM CoPilot, focusing on APIs' Fully Qualified Names (FQNs), the fundamental knowledge for effective code analysis, search and reuse.
Driven by FQNs' data distribution properties, we design a novel lightweight in-context learning on Copilot for FQN inference, which does not require code compilation as traditional methods or gradient update by recent FQN prompt-tuning.
We systematically experiment with five in-context-learning design factors to identify the best in-context learning configuration that developers can adopt in practice.
With this best configuration, we investigate the effects of amount of example prompts and FQN data properties on Copilot's FQN inference capability.
Our results confirm that Copilot stores diverse FQN knowledge and can be applied for the FQN inference due to its high inference accuracy and non-reliance on code analysis.
Based on our experience interacting with Copilot, we discuss various opportunities to improve human-CoPilot interaction in the FQN inference task.
\end{abstract}

}

\maketitle
\IEEEdisplaynontitleabstractindextext
\IEEEpeerreviewmaketitle

\input{Introduction.tex}

% \input{Paper-section/02.Background}
\input{Approach.tex}
\input{Evaluation.tex}

\input{Dis.tex}

\input{RelatedWork.tex}

\input{Conclusion.tex}
\input{samplebase.bbl}

\normalem
\balance

% \bibliography{samplebase}

% \clearpage

\par\noindent 
\parbox[t]{\linewidth}{
\noindent\parpic{\includegraphics[height=3.0in,width=1in,clip,keepaspectratio]{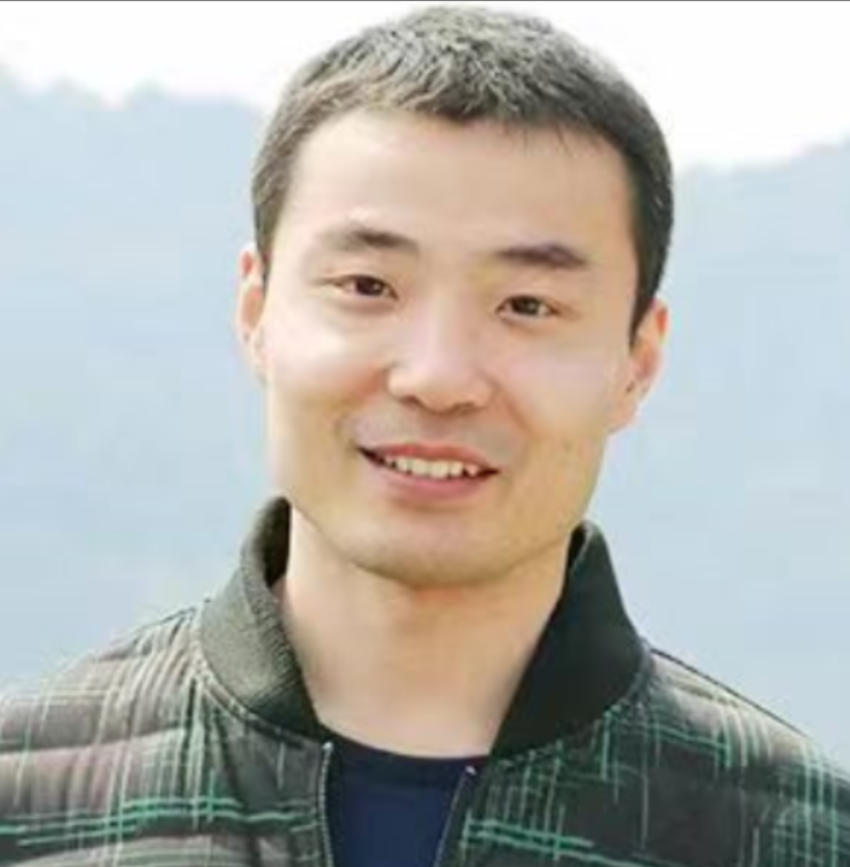}}
\noindent {\bf QING HUANG} 
received the M.S degree in computer application and technology from Nanchang University, in 2009, and the PH.D. degree in computer software and theory from Wuhan University, in 2018. He is currently an Assistant Professor with the School of Computer and Information Engineering, Jiangxi Normal University, China. His research interests include information security, software engineering and knowledge graph.}
\vspace{0.5\baselineskip}

\par\noindent 
\parbox[t]{\linewidth}{
\noindent\parpic{\includegraphics[height=3.0in,width=1in,clip,keepaspectratio]{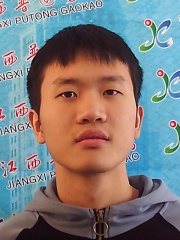}}
\noindent {\bf Dianshu Liao}\ 
is a fourth-year undergraduate student at the School of Computer and Information Engineering, Jiangxi Normal University, China. His research interests include software engineering and knowledge graph.
}
\vspace{2\baselineskip}

\par\noindent 
\parbox[t]{\linewidth}{
\noindent\parpic{\includegraphics[height=3.0in,width=1in,clip,keepaspectratio]{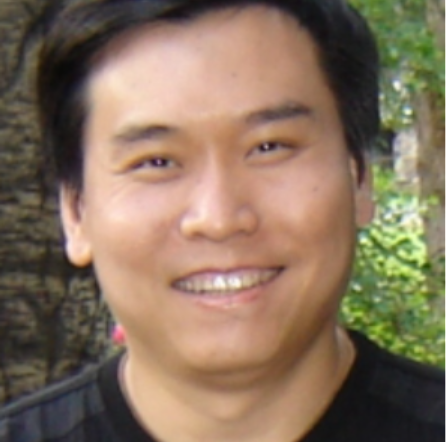}}
\noindent {\bf Zhenchang Xing}\
is a Senior Research Scientist with Data61, CSIRO, Eveleigh, NSW, Australia. In addition, he is an Associate Professor in the Research School of Computer Science, Australian National University. Previously, he was an Assistant Professor in the School of Computer Science and Engineering, Nanyang Technological University, Singapore, from 2012-2016. His main research areas are software engineering, applied data analytics, and human-computer interaction.}
\vspace{0.5\baselineskip}

\par\noindent 
\parbox[t]{\linewidth}{
\noindent\parpic{\includegraphics[height=3.0in,width=1in,clip,keepaspectratio]{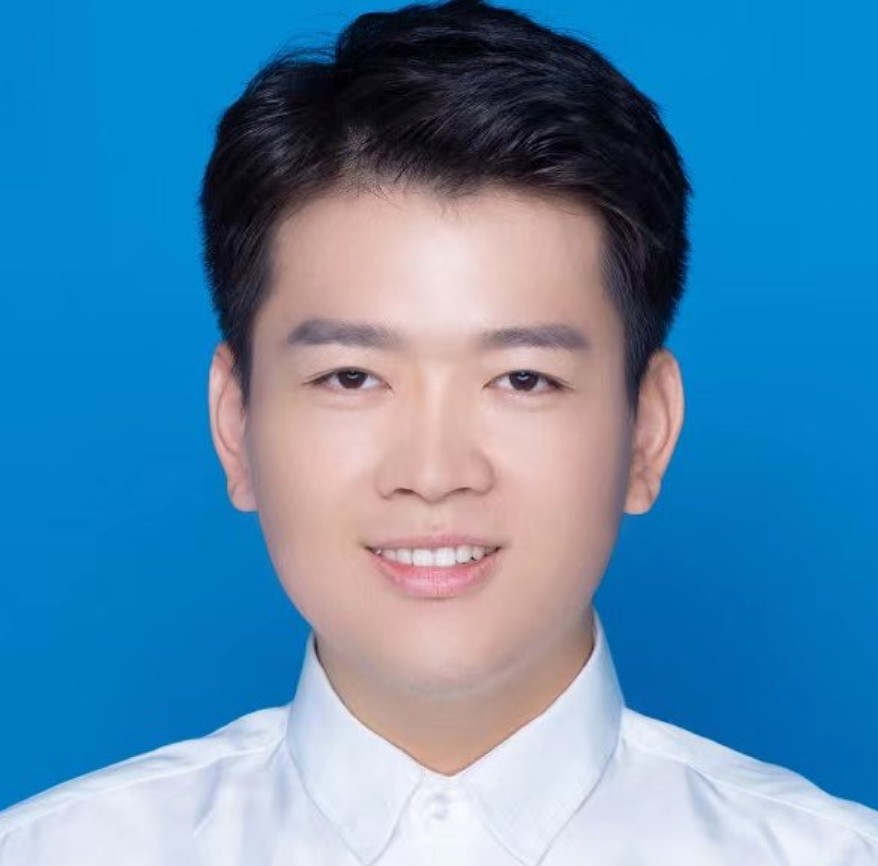}}
\noindent {\bf Zhiqiang Yuan}\
is a second-year graduate student in the School of Computer and Information Engineering, Jiangxi Normal University. His research interests are software engineering and knowledge graph.
}
\vspace{1\baselineskip}

\par\noindent 
\parbox[t]{\linewidth}{
\noindent\parpic{\includegraphics[height=3.0in,width=1in,clip,keepaspectratio]{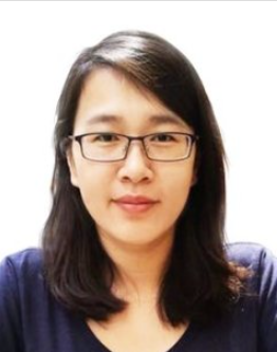}}
\noindent {\bf Qinghua Lu}\
is a Senior Research Scientist with Data61, CSIRO, Eveleigh, NSW, Australia. Before joining Data61, she was an Associate professor at China University of Petroleum, and she worked as a researcher at National Information and Communications Technology Australia. She has published more than 100 academic papers in international journals and conferences. Her research interests include the software architecture, blockchain, software engineering for AI, and AI ethics.}
% \vspace{1\baselineskip}

\par\noindent 
\parbox[t]{\linewidth}{
\noindent\parpic{\includegraphics[height=3.0in,width=1in,clip,keepaspectratio]{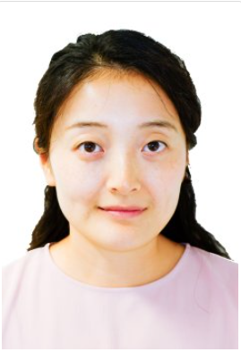}}
\noindent {\bf Xiwei Xu}\
is a Senior Research Scientist with Architecture\& Analytics Platforms Team, Data61, CSIRO. She is also a Conjoint Lecturer with UNSW. She started working on blockchain since 2015. She is doing research on blockchain from software architecture perspective, for example, tradeoff analysis, and decision making and evaluation framework. Her main research interest is software architecture. She also does research in the areas of service computing, business process, and cloud computing and dependability.}
% \vspace{0.5\baselineskip}

\par\noindent 
\parbox[t]{\linewidth}{
\noindent\parpic{\includegraphics[height=3.0in,width=1in,clip,keepaspectratio]{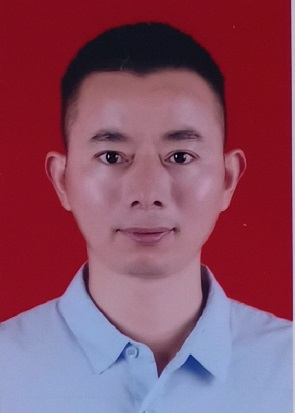}}
\noindent {\bf Jiaxing Lu}\
received his Master's degree from Jiangxi Normal University, China, in 2004. He is currently a Lecturer in the School of Computer and Information Engineering of Jiangxi Normal University. His research interest is mainly in the areas of algorithms of artificial intelligence and computing theory. He has participated in multiple projects and published several research papers in scholarly journals in the above research areas.
}
% \vspace{0.5\baselineskip}

\end{document}

%% file: Introduction.tex
% \vspace{-2mm}
\section{INTRODUCTION}

% \vspace{-1mm}

% \zc{Try to revise according to this intro skeleton. Expand this paragraph with more details. Move your intro content to appropriate places in the skeleton. Try to control the length of intro in 1.5 page.}

%Problem context. You should start with what SE cares about, not LMs. 开门见山，直奔主题，这个工作是关于SE的什么事情。
Software engineering (SE) is knowledge-intensive.
The knowledge includes, but not limited to:
1) factual knowledge, such as the API's fully-qualified name (FQN), user credentials, text resources. 
2) structural knowledge, such as Abstract Syntax Tree (AST) and control flow graph (CFG), reflecting the code's syntax and execution flow.
%, they reflect the code's syntactic and execution logic.
3) semantic knowledge, such as code patterns, code semantics and API constraints, usually lead to bugs or vulnerabilities if they are overlooked.

% It includes:
% factual knowledge (e.g., what is the fully-qualified name of an API), 
% Structural or syntactic knowledge (e.g., Abstract Syntax Tree), and Semantic knowledge (e.g., bugs or vulnerabilities).

%A brief summary of NLP advances and its relation to SE knowledge and what has been studied. Basically many studies on syntactic and semantic knowledge in code models. give some examples.
In the Natural Language Processing (NLP) community, many studies~\cite{Peters2018DeepCW, Radford2018ImprovingLU,Devlin2019BERTPO,Yang2019XLNetGA,Liu2019RoBERTaAR,Brown2020LanguageMA} show that the pre-trained language models (PLMs) have powerful capabilities for capturing linguistic and semantic knowledge in text.
These capabilities enable PLMs to significantly improve the state-of-the-art in NLP tasks.
In the SE community, code naturalness shows that code can be understood and manipulated in the same way as natural language text~\cite{Devanbu2012OnTN, Allamanis2018ASO}.
This spawn a series of pre-trained code models (PCMs), e.g., CodeBERT~\cite{Feng2020CodeBERTAP}, CodeT5~\cite{wang2021codet5}, CoPilot~\cite{GithubCopilot}, modeling code simply as text.
Studies have shown that syntactic or semantic knowledge packed in the PCMs can be transferred to the downstream SE tasks and benefit these tasks~\cite{Wan2022WhatDT, Buratti2020ExploringSN, yuan2022deep, Wang2022BridgingPM, Karmakar2021WhatDP, wang2022no, Qiu2020PretrainedMF, Han2021PreTrainedMP}.
Giant PCMs (e.g., CoPilot~\cite{GithubCopilot}) start coming into the everyday software development practices.
%For example, two recent studies~\cite{Wan2022WhatDT, Buratti2020ExploringSN} confirm that PCMs (e.g., CodeBERT) preserve the syntax structure of code, and Yuan et al.~\cite{yuan2022deep} utilize CodeBERT to detect the vulnerabilities in the source code.

% ??? studied how to use ??? to learn semantic and syntactic features in code and use them to predict vulnerabilities.

% @article{yuan2022deep,
%   title={Deep Neural Embedding for Software Vulnerability Discovery: Comparison and Optimization},
%   author={Yuan, Xue and Lin, Guanjun and Tai, Yonghang and Zhang, Jun},
%   journal={Security and Communication Networks},
%   volume={2022},
%   year={2022},
%   publisher={Hindawi}
% }

% Dam et al.~\cite{17vulnerabilityprediction} studied how to use \zc{a long and short-term memory model ??LSTM is not pre-trained code model. This work is irrelevant to the point we want to explain in this paragraph! There are so many works using PCMs for vulnerability prediction. Why not use the PCM one?} to learn semantic and syntactic features in code and use them to predict vulnerabilities.

% Quickly to our study focus: SE factual knowledge, but no study on this type of SE knowledge. what specific SE factual knowledge we focus on in this work and why it is important to study.
However, there has been no work on probing and using SE factual knowledge in the PCMs.
In this work, we study a particular type of SE factual knowledge, i.e., API's FQNs.
An API FQN identifies which class, function or field a simple name refers to in a code context.
It is a fundamental knowledge for program analysis~\cite{godefroid2008automating, Zhou2019DevignEV, Ren2020APIMisuseDD,TypeInferenceforC, Zhang2018AreCE, Piccolboni2021CRYLOGGERDC} and code search~\cite{Maji2021DCoMAD, thummalapenta2007parseweb}.
%is very important in software engineering~\cite{Gupta2020JCoffeeUC,thummalapenta2007parseweb}.
%It not only helps program analysis techniques make full use of the knowledge in the code~\cite{godefroid2008automating,lattner2004llvm}, but also helps code search engines~\cite{Maji2021DCoMAD, thummalapenta2007parseweb} recommend relevant code examples more accurately.
%Furthermore, the FQN of an API used in the code contains accurate API type information, which can help vulnerability analysis tools~\cite{Zhou2019DevignEV, Ren2020APIMisuseDD,TypeInferenceforC} to detect API misuses or even malicious behaviors~\cite{Zhang2018AreCE, Piccolboni2021CRYLOGGERDC}.
However, the FQNs of simple names in partial code (see Fig.~\ref{fig:approach figure}) usually cannot be resolved which hinders the partial code reuse and analysis~\cite{rahman2018evaluating,gopstein2018prevalence,yin2018learning}.  
Existing work~\cite{Baker, Saifullah2019LearningFE, Su2018SNRConstrainedHF} for FQN inference in partial code depends on a symbolic knowledge base of APIs and their usage and partial program analysis~\cite{Dagenais2008EnablingSA}.
However, building such symbolic knowledge bases requires project compilation, and suffers from out-of-vocabulary (OOV) issues~\cite{Huang2022PrompttunedCL}.
Our previous work~\cite{Huang2022PrompttunedCL} proposed to treat code as text and infer FQNs as a text fill-in-blank task based on an FQN-prompt-tuned CodeBERT (see an illustration in Fig.~\ref{fig:approach figure}), showing the promise of PCMs for modeling FQN knowledge in code.

%Introduce giant model and its in-context learning ability which inspires our work 
Supervised fine-tuning~\cite{Wan2022WhatDT,Huang2022PrompttunedCL,shi2021overcoming} updates the model weights.
Although the fine-tuned PLMs perform well in the downstream tasks, it does not mean that the vanilla PLMs without fine-tuning effectively store relevant knowledge.
In fact, our study (see RQ4 in Section~\ref{sec:RQ4 compare with mlm}) shows that CodeBERT (125M parameters)~\cite{Feng2020CodeBERTAP} (the backbone PCM fine-tuned in~\cite{Huang2022PrompttunedCL}) performs poorly on FQN inference without fine-tuning.
A recent study by Google~\cite{Wei2022EmergentAO} shows that model size matters and emergent abilities will not appear until a critical threshold of scale is reached.
Recent NLP studies~\cite{Brown2020LanguageMA, Petroni2019LanguageMA,Heinzerling2021LanguageMA} demonstrate that giant PLMs (e.g., GPT3 with 175B parameters) have strong ability to store factual and commonsense knowledge as neural knowledge base and the stored knowledge can be accurately recalled through in-context learning (see Fig.~\ref{fig:approach figure} for the illustration of in-context learning).
In these studies, giant PLMs are frozen (i.e., without updating model parameters) such that people can understand to what extent certain factual knowledge is present in the intact models.
%, rather than manifested in the downstream tasks.

% \begin{figure}%[H]
%     \centering
%     \includegraphics[width=0.47\textwidth]{picture/FQN_Usage_Distribution.png}
%     \caption{The API Usage Distribution for 229,798 unique FQNs. The horizontal coordinate is the unique FQN and the vertical coordinate is the number of times the FQN has been used. \zc{The blue section contains the top 80 percent of FQN used times. ??maybe blue for $\geq 2$}\liao{Because of the page limitation, I think this figure can be deleted.}} 
%     \label{fig:the power law distribution of FQN usage}
%     % \vspace{-0.5cm}
% \end{figure}

%Now our study of FQN data distribution
A recent study~\cite{Chan2022DataDP} identifies three data distribution properties (temporally bursty, Zipfian distribution, and dynamic meaning) empowers the in-context learning ability of frozen giant PLMs.
FQN data manifests these three properties.
First, the temporally bursty is reflected in API evolution~\cite{xing2007api,lamothe2021systematic,Li2020CDACD,Robbes2012HowDD}, during which new APIs are introduced and existing APIs are removed, changed or deprecated.
Furthermore, new libraries emerge and existing ones become obsolete~\cite{xing2007api}.
%For example, Robbes et al.~\cite{Robbes2012HowDD} shows that the usage of \textit{FillInTheBlank} gradually increased over time and then suddenly decreased as developers switched to \textit{UIManager}, causing a sudden spike in \textit{UIManager} usage.
Second, the Zipfian (i.e. power law) distribution is reflected in the FQN lengths and usage.
In our experimental dataset of six libraries (Android SDK, JDK, Hibernate, GWT, Xstream, Joda Time), about 12\% (or 27\%) FQNs have been used more than 10k (or 1k) times respectively, followed by a long tail of infrequently used FQNs (see Table~\ref{Table:accuracy of different FQN data distribution properties}).
%\liao{The percentage in Table~\ref{Table:accuracy of different FQN data distribution properties} is calculated based on the sampled dataset, not the original six github library. Does the percentage in the experiment result need to be calculated based on the original six library dataset?}. 
Third, the dynamic meaning is reflected in the 1:N and N:1 cardinalities between simple names and FQNs (i.e., polysemy and synonymy ambiguity in NLP).
%, i.e., the same FQN can be declared as different variable names, and the same simple name corresponds to the different FQNs.
For example,  all variables \textit{reader}, \textit{br} and \textit{buffRead} can be of the type \textit{java.io.BufferedReader}.
The simple name \textit{Date} can be \textit{java.util.Date}, \textit{java.sql.Date} or \textit{sun.util.calendar.Gregorian.Date}.

%Our lightweight FQN inference by in-context learning 
We hypothesize that frozen giant PCMs (e.g., CoPilot~\cite{GithubCopilot} extended from GPT3) can serve as a neural knowledge base of SE factual knowledge (e.g., FQNs) because they are pre-trained on a giant corpus of source code.
To validate this hypothesis, we conduct a series of experiments, focusing on two questions:
% 1) to what extent frozen giant code model (Copilot in this study) store FQN knowledge?
% 2) can we retrieve FQN knowledge in Copilot without fine-tuning and by what means?
1) how much FQN knowledge is packed in a frozen giant PCM (CoPilot in this study)?
2) how can we effectively retrieve the FQN knowledge in the frozen CoPilot?
%the frozen giant pre-trained code model by in-context learning?
Inspired by the alignment of FQN data distribution properties with the findings in~\cite{Chan2022DataDP}, we set up in-context learning tasks on CoPilot for inferring FQNs for cannot-be-resolved simple names in partial code snippets (see Fig.~\ref{fig:approach figure}).
We experiment a wide range of learning configurations:
zero/one/few-shot learning and in-context learning design factors (code context, task description, prompt template, example prompt order and identifier format).
We analyze the CoPilot's capability in inferring FQNs with diverse data distribution properties (FQN lengths,  FQN usage times, simplename-FQN and FQN-simplename cardinalities).
%As a result, we identify an effective prompt design and confirm that Copilot can effectively support FQN inference in partial code in practice. 

As CoPilot can only be manually invoked, we sample a set of 1,440 representative and diverse methods from the Github dataset of six libraries in~\cite{Huang2022PrompttunedCL}.
These methods use 4,697 distinct APIs from 850 packages.
%Our experiments are done on an FQN dataset from 1,440 methods which use 8,258 APIs from 1,440 packages.
%These methods are sampled from the Github dataset of six libraries in~\cite{Huang2022PrompttunedCL}.
%The average cosine similarity of each method pairs is 0.48, and the average Jaccard similarity between FQN sets of each two methods is 0.027.
%Comparison with Zhiqing's ASE
%Compared with Yuan et al.~\cite{Huang2022PrompttunedCL}, 
Different from supervised FQN-prompt tuning~\cite{Huang2022PrompttunedCL}, our FQN inference stands on the shoulder of frozen giant CoPilot.
With the lightweight in-context learning, CoPilot can be quickly adapted to the FQN inference task with zero or a few examples of simplename-FQN mappings.
On the Stack Overflow dataset (496 code snippets using 791 distinct APIs from 66 packages), our method achieves very close or better inference accuracy than supervised prompt tuning of CodeBERT with 11,776 library source files in~\cite{Huang2022PrompttunedCL}.
Our study also unveils several interesting interactions with CoPilot, including micro-level sensitivity, human-CoPilot colearning and priori knowledge for in-context learning.
These interaction experiences call for further studies on human-PCM collaboration design to effectively integrate human intelligence and AI in software engineering tasks.

%\zc{will finalize later ...}
The main contributions of this paper are as follows:

\begin{itemize}
% \vspace{-1mm}
    \item Conceptually, we conduct the first systematic study on a fundamental SE factual knowledge (FQN) in the state-of-the-art giant PCM (CoPilot). Our methodology can be extended to other SE factual knowledge in giant PCMs (e.g., privacy and proprietary information in code).
    %in parthe first systematic work to investigate the capabilities and limitations of a large pre-trained code language model for inferring factual software engineering knowledge (FQN knowledge in particular)

    \item Technically, we design the first lightweight in-context learning-based method for FQN inference, standing on the shoulder of frozen giant PCM.
    Our method removes the reliance on code compilation and special model tuning and deployment. 
    Developers can easily adopt our method when reusing and parsing partial code by writing a few lines of code comments demonstrating the FQN inference task to CoPilot and then requesting the CoPilot's completion.
    %built on the back of a frozen commercial giant code language model (as opposed to fine-tuned model based on masking strategy~\cite{Huang2022PrompttunedCL}, which is based on a small research prototype MLM and requires fine-tuning MLM parameters). 
    
    \item Empirically, we systematically experiment a wide range of in-context learning configurations. Our results reveal the extent and characteristics of FQN knowledge stored in CoPilot and confirm the practicality of using CoPilot for FQN inference in partial code. We also identify effective in-context learning configurations for different priori FQN knowledge and data properties.
    %Extensive empirical studies of in-context learning settings, including prompt engineering, zero vs one(example not in code) one vs few vs few(leave one out), FQN data distribution properties, compare with baseline. 
    Our data package can be found here~\footnote{\href{https://github.com/SE-qinghuang/SE-Factual-Knowledge-in-Frozen-Giant-Code-Model}{https://github.com/SE-qinghuang/SE-Factual-Knowledge-in-Frozen-Giant-Code-Model}}.
    Code will be released upon paper acceptance.
    %\item partial code reuse, practicality ... convenient to deploy ...
    
\end{itemize}

%% file: Approach.tex
% \vspace{-1mm}
\section{In-Context Learning for FQN Inference}

\begin{figure*}%[H]
    \centering
    \includegraphics[width=1\textwidth]{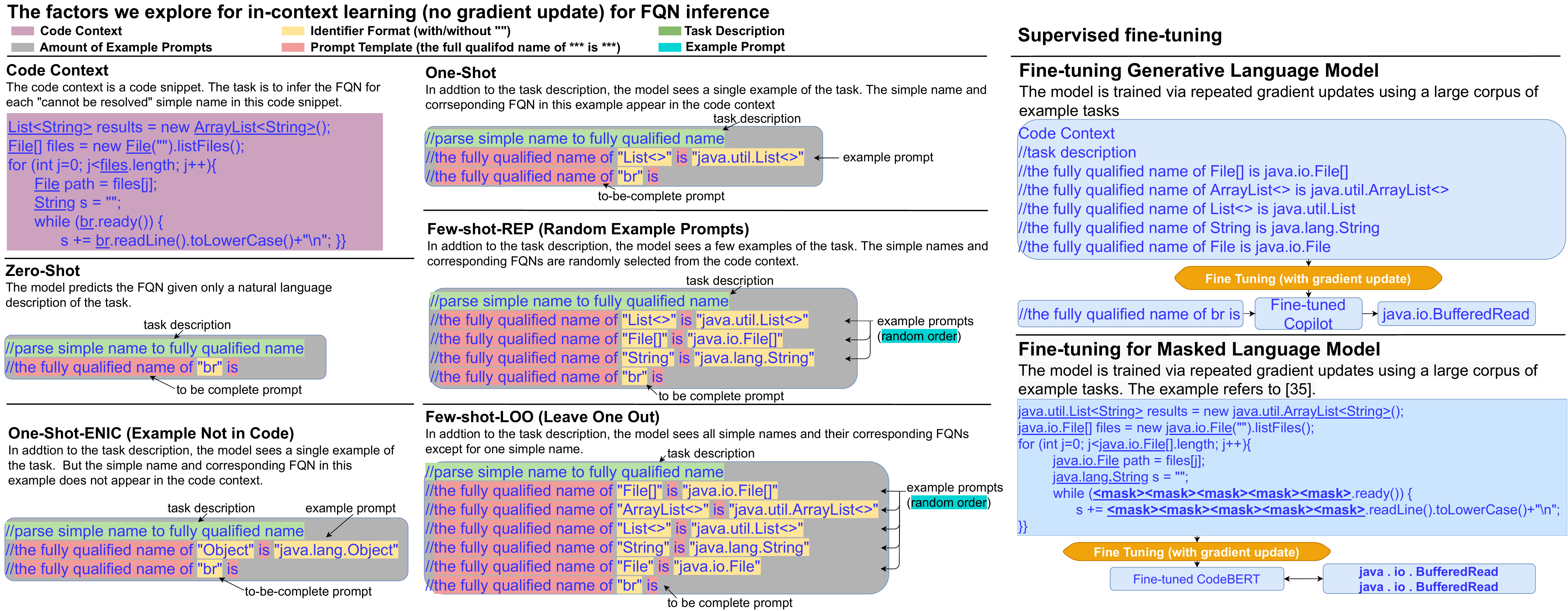}
    % \vspace{-8mm}
    \caption{In-Context Learning for FQN Inference. The bold text is the explanation, not part of task input. The model's task input parts are highlighted, such as the code context in purple block and the prompts in gray block. A task input concatenates a purple block and a gray block. 
    % \zc{Shot names need update like One-Shot-ENIC (Example Not in Code)}
    % \zc{1) Add reference to our ASE2022 to MLM Fine-tuning example. \liao{Done.But i want to keep this comment until all the context is finalized, beacuse changes in the references will change this [32].}}
    } 
    \label{fig:approach figure}
    % \vspace{-4mm}
\end{figure*}

%\zc{Introduce in-context learning basis, describe basic configuration. Illustrate basic configuration and highlight six factors in a figure like GPT3. Contrast in-context learning to supervised fine-tuning.}

%\zc{Maybe some concise background about pre-trained LLM, prompt learning, in-context learning before basic configuration?}

%in-context learning的基本概念，它在nlp中怎么setup的
%in-context learning在type inference中的六个factors
%in-context learning在type inference中的六个factors怎么

%We briefly review relevant background knowledge and explain our design of in-context learning for FQN inference, and how we apply it to Copilot.
%how we apply in-context learning on frozen giant PCM (i.e., Copilot~\cite{GithubCopilot}) for API FQN inference.
We design a lightweight, easy-to-deploy in-context learning method for FQN inference in partial code.
This method allows us to probe the FQN knowledge in the frozen giant PCMs.

% 2) supervised fine-tuning: heterogeneous downstream task fine tuning (applicable even without pre-trained lm); prompt convert downstream tasks into pre-training objectives (zero - direct reuse; one/few shot good capability but can be fine-tuned with large data)

% \vspace{-2mm}
\subsection{Supervised Fine-Tuning vs. In-Context Learning}\label{sec:backgroundknowledge}
% \vspace{-1mm}

\subsubsection{Supervised Fine-Tuning}
PCMs have been adopted in many downstream SE tasks through supervised fine-tuning~\cite{Feng2020CodeBERTAP,zhang2020graph,chen2021evaluating}.
Fine tuning adapts task-agnostic PLMs to downstream tasks by gradient-updating the weights of a PLM on a supervised dataset specific to the downstream task.
%In this setting, the pre-trained model weights are used to initialize a task-specific model which is then further fine-tuned.
%task-specific fine-tuning of the PLMs on the task-specific occurs mainly in the downstream heterogeneous.
Downstream tasks are generally heterogeneous from the PLM pretraining.
To alleviate this heterogeneity, prompt tuning has been proposed~\cite{liu2021pre, Brown2020LanguageMA, schick2020automatically}.
%to bridge the gap of learning objectives in model pre-training and fine-tuning.
Prompts are templates that convert downstream tasks into the form of pre-training tasks which makes pre-training and prompt tuning share homogeneous learning objective.
``Pre-training, prompt-tuning and prediction'' has become a new NLP paradigm and has demonstrated strong capability in zero-, one- or few-shot scenarios in code summarization~\cite{sun2022extractive}, fault prediction~\cite{rathore2019study}, code translation~\cite{lachaux2020unsupervised} and requirement classification~\cite{wang2022no}.
Recently, we.~\cite{Huang2022PrompttunedCL} applied FQN-prompt-tuning to CodeBERT for FQN inference.
However, as fine-tuning updates model weights, we cannot know how much FQN knowledge a frozen PLM captures.
%Because prompt learning better integrates pre-training and downstream tasks and brings NLP tasks closer to human logic and habits.

% achieved promising results on some other few-class
% General, there has a gap between the fine-tuning task and the pre-trained model objective,  i.e., heterogeneous
% For example, xxx fine-tuning the xxx to the xxx.
% xxx proposed prompt, and through the prompt to convert the downstrame tasks with pre-training objective, i.e....
% xxx shows that through prompt tuning, the model preformance well in the zero-few shot.
% % L --> yigou --fine
% % promp --> 一致起来 --> fine- --> few/one-shot ability

% the task-specific learning
% occurs mainly in the downstream heterogeneous neural network.
% We refer to this paradigm as outsourcing upgrade. In contrast, we
% adopt a completely new paradigm “pre-train, prompt and predict”
% which is a self-upgrade of pre-trained code MLM, in the sense that
% model pre-training and prompt learning are homogeneous

% 3) in-context learing: only applicable to pretrained LM when convert downstream tasks into pre-training objectives (zero usually still have task description, if no taskdesc, direct reuse plm; one/few shot good capability and context windwos limits few shot)
\subsubsection{In-context Learning}

In-context learning is a form of prompt learning without gradient update of PLMs so that we can learn to what extent a frozen PLM captures certain knowledge.
In-context learning uses the text input to a PLM to specify the downstream task: the PLM is conditioned on a task description and/or a few task demonstrations (i.e., example prompts) and is then asked to complete further instances of the task (i.e., to-be-complete prompt) by generating what comes next.
%As show in the 
Fig.~\ref{fig:approach figure} presents our task input for FQN inference.
%, the task description in our work can be ``parse simple name to fully qualified name''.
Although a PCM may have seen many simple names and FQNs during pre-training, this form of the FQN inference task has never been seen during pre-training. 
By analyzing the capability of a PCM in generating the FQN for the given simple name in a code context in the in-context learning setting, we probe the FQN knowledge that this PCM captures.

% \vspace{-3mm}
\subsection{Our In-Context Learning Design}\label{sec:Our In-Context Learning Design}
% \vspace{-1mm}
% \zc{reshuffle the content as two main perspectives: 1) amount of example prompts. 2) prompt engineering (the other five factors). As the five shot settings are always there for all experiments, I think we should treat it as a stand-alone perspective, higher than the other five factors.}

To develop a comprehensive understanding of the FQN knowledge in a frozen, giant PCM, we consider two main perspectives in designing in-context learning tasks.
% \vspace{-3mm}
%Next, we introduce the factors in designing in-context learning task for API FQN Inference.
%Based on In-context learning concept, we set up several different learning scenarios, i.e., zero-shot, one-shot and few-shot learning. 
%Few-shot learning is given a few demonstrations of the task at inference time as conditioning, but no weight updates are allowed.
%As shown in Figure~\ref{fig:approach figure}, for a typical dataset an example has a context and a desired completion (for example
%an type inference), and few-shot works by giving K examples of context and completion, and then one final example of context, with the model expected to provide the completion.
%As such, K=0 is zero-shot learning, and K=1 is one-shot learning.

%\subsubsection{Six Design Factors for In-Context Learning for API FQN inference}

%There are six types of factors for in-context learning for FQN inference.
%These six factors can generate a prompt that activates the pre-trained model to recover the FQN knowledge and accomplish the FQN inference task.

\subsubsection{Amount of Example Prompts}\label{sec:amount of example prompts}
Example prompts are the most important part for in-context learning. 
As shown in Fig.~\ref{fig:approach figure}, example prompts follow the task description and appear before the to-be-complete prompt.
They not only inform the model priori-known FQNs but also ``teach'' the model how it should complete the task unseen during pre-training.
Given a code snippet, the developer may know the FQNs of some simple names.
Priori-known FQNs would help to infer other unknown FQNs.
To simulate different levels of priori-known FQNs in the FQN inference task, we design five different shot settings (zero-shot, two one-shot and two few-shot).

%This known information can be used as examples to activate the model to recall the FQN knowledge and thus better perform the task of FQN inference.
%These examples can be used as part of the prompt to help the model quickly adapt to new task scenarios.
%According to the number of known examples, the shot type can be classified into five types.

The \textbf{zero-shot} provides no example prompts.
This simulates the situation when the developer does not know any FQN (no matter for the simple names in the code context or any other FQNs).
As shown in \textit{Zero-Shot} in Fig.~\ref{fig:approach figure}, the model predicts the FQN of the simple name given only the task description.
Given this minimum task input, the model may not clearly know what task to solve and thus generate some irrelevant results (e.g., generating a piece of code instead of an FQN, generating the value of a variable instead of its FQN).

The \textbf{one-shot-ENIC (example not in code context)} provides one example prompt, but the simple name and corresponding FQN does not appear in the code context.
This setting assumes that even if the developer does not know the FQNs of any simple names in the code context, he or she may still know some general FQNs.
%This type of prompt is constructed when the programmer does not know any simple name corresponding FQN that appears in the code context, but knows one simple name and its corresponding FQN outside the code context.
As shown in \textit{One-Shot-ENIC (Example Not in code context)} in Fig.~\ref{fig:approach figure}, the example prompt shows the simple name \textit{Object} and its corresponding FQN \textit{java.lang.Object}, even though \textit{Object} does not appear in the code context.
Although this example prompt does not provide any priori-known FQNs relevant to the code context, it still demonstrates the task completion format, which may reduce the chance of generating the text irrelevant to FQN.
%This type of prompt can help the model to know the answer format of the questions and thus generate a more standardized result.
However, this not-in-code-context simplename-FQN example may cause the model to misunderstand the scope of FQN inference and generate an FQN irrelevant to the code context.
%i.e., the simple name whose FQN to be inferred does not necessarily belong to the provided code context.

The \textbf{one-shot} provides one example prompt for a simple name (selected randomly or by FQN usage times) appearing in the code context.
%This prompt is constructed when the programmer knows a simple name and its corresponding FQN  appears in the code context.
As shown in \textit{One-Shot} in Fig.~\ref{fig:approach figure}, the prompt shows the simple name \textit{List$<>$} and its corresponding FQN \textit{java.util.List$<>$}, and the simple name \textit{List$<>$} appears in the code context.
This setting is the lower bound of priori-known FQN in the code context and demonstrates the task completion format.
%know the format of the answer to the question and the scope of the answer is the FQN of a simple name in the code context.
However, providing only one example may not be sufficient to adapt the model to the FQN inference task.
%other FQNs in the code context.

The \textbf{few-shot-REP (random example prompts)} provides the example prompts for 2 to $n-2$ randomly selected simple names in the code context.
$n$ is the number of all unique cannot-be-resolved simple names.
In the example of \textit{Few-Shot-REP (Random Example Prompts)} in Fig.~\ref{fig:approach figure}, there are three example prompts, each for one simple name in the code context.
This simulates the situations when the developer knows the FQNs for some simple names but not others.
The more example prompts the model sees, the more likely it can adapt to specific code contexts and infer relevant FQNs in the correct format.
%The three examples are randomly selected from all the examples that appear in the code context.
%This prompt usually activates the model to recall the FQN knowledge effectively, because the more the number of examples, the more the model can adapt the task and determine the answer format, the answer scope, and the answer content.

The \textbf{few-shot-LOO (leave one out)} provides the example prompts for all $n-1$ simple names in the code context, except for one simple name left out for inference, as illustrated in the example of \textit{Few-shot-LOO (Leave One Out)} in Fig.~\ref{fig:approach figure}.
This setting is the upper bound of priori-known FQNs, which would maximize the model's capability in inferring the FQN for the left-out simple name.
This upper bound allows us to estimate the extent of FQN knowledge in a giant PCM.
%This prompt is constructed when the programmer knows all the simple names and corresponding FQNs except for the predicted one simple name.
%As shown in the part of \textit{few-shot (leave one out)} in Fig.~\ref{fig:approach figure}, the prompt shows all examples in which all simple names appear in code context except the predicted simple name.
%Usually, this type of prompt can maximize the model's activation to recall the FQN knowledge and effectively complete the FQN inference task.

\subsubsection{Prompt Engineering}\label{sec:promptengineering}

In addition to task demonstrations by example prompts, our in-context learning considers the following five factors when preparing the task input.

\textbf{Code Context.}
The code context is a code snippet (the purple block in Fig.~\ref{fig:approach figure}).
We put the code context at the beginning of the task input, which specifies the context where the FQN inference occurs.
If the code context is not given, we essentially probe the preferred FQN that the model memorizes for a simple name from pre-training.
%As this code is incomplete, underlined simple names cannot be resolved.

% \zc{need to explan how to get List<> from List<String>, get File[] from File[100] or get List() from List(...). also explain we treat List, List<> and List[], List() as different simple names because we want to problem the model's capability in understanding different program elements (type, generic type, array type, constructor)}

%Each simple name forms a context with the rest of the code snippet, thus helping the model to determine the probability of the simple name in the presence of such a code context.

\textbf{Task Description.}
The task description (the green block in Fig.~\ref{fig:approach figure}) follows the code context and appears before the example prompts.
It tells the model what the task is about.
We consider three types of task description: no description, concise (i.e., ``type inference''), or verbose (i.e., ``parse simple name to fully qualified name'').
%The position of the task description is usually in the position under the code context and before the examples, such as the "parse simple name to fully qualified name in five shot prompts in Fig.~\ref{fig:approach figure}.

\textbf{Prompt Template.}
All example prompts and to-be-complete prompt use the same prompt template so that the model can learn from the example prompts how to complete the to-be-complete prompt.
%The answer format is a template prompt that helps the model generate the result (e.g., the fully qualified name).
We experiment two types of prompt templates: description or symbol.
The description template is ``the fully qualified name of \textit{simplename} is \textit{FQN}'', as shown in Fig.~\ref{fig:approach figure}.
%In general, the answer format can be either symbol or description type.
The symbol template uses a symbol ($\rightarrow$ in our experiments) to indicate mapping a simple name to an FQN (e.g., \textit{File} $\rightarrow$ \textit{java.io.File}).
%The answer format of description type usually prompts the model to predict based on the natural language, such as the answer format with the pink block in Fig.~\ref{fig:approach figure} \textit{the fully qualified name of simplename is FQN}.

\textbf{Order of Example Prompts.}
To understand if the order of example prompts affects the model's learning, we design three orders: random, frequent first, and infrequent first.
We count the FQN usage times in all the methods in our dataset.
Frequent-first order means the example prompts of the more frequently used FQNs appear before the less frequently used FQNs.
Infrequent-first order is the opposite to Frequent-first order.
Random order means randomly ordering example prompts without considering the FQN usage times. 

%The level to activate the model to recall factual knowledge is affected by the order of the examples provided in some degree.
%The examples can be ordered from frequent FQNs to infrequent FQNs, infrequent FQNs to frequent FQNs, or randomly.
%The frequency of an FQN can usually be measured by the number of method calls it receives.
%The more methods that call it, the more frequently the FQN appears.

\textbf{Identifier Format.}
Annotating a word in a sentence can help the model distinguish it from other words.
We experiment the identifier format with and without special annotations.
Without annotations means the words in the prompts are treated equally.
With annotations, we add \textit{``''} to simple names and FQNs in the prompts, for example, \textit{the fully qualified name of ``File'' is ``java.io.File''} in Fig.~\ref{fig:approach figure}.
%to help the model assign more weight to them.
%For example, the simplename is remarked as "simplename" and the FQN is remarked as "FQN" in the prompt in Fig.~\ref{fig:approach figure}.

% \vspace{-1mm}
\subsection{FQN Inference by In-Context Learning on CoPilot}
\label{sec:application of incontext learning}
% \vspace{-1mm}

%Six major factors influence the effect of activating FQN information utilizing the in-context learning technique.
%They are code context, shot-type, task description, answer format, example order, identifier type.
%We must find some theoretically feasible components and investigate the influence of one of them on the model's estimated effect while keeping the other component variables constant.
In this study, we probe the FQN knowledge in CoPilot with 175B parameters.
We choose CoPilot as it has commercial product quality and provides the IDE plugin to invoke the model.
We use giant PCM rather than small models like CodeBERT~\cite{Feng2020CodeBERTAP} as the study~\cite{Wei2022EmergentAO} shows giant models exhibit emergent abilities that small/medium models do not have.
To apply in-context learning on CoPilot for FQN inference, developers can write a task input comprising one code snippet as text (if provided), followed by one task description, example prompts (if any) (one per line) and one to-be-complete prompt at the end, as shown in Fig.~\ref{fig:approach figure}.
Each part starts on a new line.
Task description and prompts starts with // as code comments.
The entire task input is plain text, and CoPilot can be invoked on the to-be-complete prompt line to complete it.
%We get these pieces of content through the following five steps.

%shows, we set the basic configurations for the six component variables.

%\subsubsection{Application of In-Context Learning}

A to-be-complete prompt is to infer the FQN for a cannot-be-resolved simple name in the code context.
As in~\cite{Huang2022PrompttunedCL}, we consider three types of cannot-be-resolved simple names:
1) data type of variable declaration (e.g., ``File'' in Fig.~\ref{fig:approach figure});
2) type name of class instantiation and array creation (e.g., ``List$<$String$>$'');
3) the object or the type on which a method is invoked or a field is accessed (e.g., ``br'').
In addition, we consider the method/field name of this method invocation or field access (e.g., readLine()) to test the CoPilot's inference capability for methods and fields.
However, we do not consider the chained method calls or field accesses (e.g., ``br.readLine().toLowerCase()'').
Once ``br'' or ``readLine()'' is resolved to an FQN, the receiving type of ``toLowerCase()'' can be obtained from the return type of ``br.readLine()''.
%We populate the identified type inference points into a custom prompt template (i.e., ``the full qualified name of *** is ***'' in Fig.~\ref{fig:approach figure}). }

As shown in Fig.~\ref{fig:approach figure}, each prompt (example or to-be-complete) contains only one simplename-FQN mapping.
%Note that each prompt contains one API type that needs to be inferred.
For example, for ``List$<$String$>$'', there will be two separate prompts, one for ``List$<>$'' and the other for ``String''.
%we use two prompts to infer FQN for ``List$<$$>$'' and ``String'' independently.
Furthermore, we treat simple names with same base but different forms (e.g., ``List'', ``List$<>$'', ``List[]'', and ``List()'') as different simple names, because we want to probe the CoPilot's capability in understanding different program syntax (i.e., general type, generic type, array type, constructor).

%% file: Evaluation.tex
\section{Experiments Setup}

Our experiments have two-fold objectives:
1) investigate how much FQN knowledge is packed in a frozen giant PCM like CoPilot; and
2) identify the effective ways to retrieve the FQN knowledge in CoPilot by in-context learning.
%To evaluate the effectiveness of the in-context learning-based approach for inferring factual software engineering knowledge (FQN), we stand on the shoulders of the frozen giant code language model (e.g., CoPilot) to solve the FQN inference task. 
To achieve these objectives, we conduct a series of experiments to investigate the four research questions:

\begin{itemize}

    \item RQ1 - How sensitive is the FQN inference on CoPilot to five prompt engineering factors?

    \item RQ2 - How do amount of example prompts (zero/one/few-shot) affect the FQN inference on CoPilot?

    \item RQ3 - How do FQN data distribution properties affect the FQN inference on CoPilot?

    %\item \zc{RQ3 - What is the Copilot's capability in inferring FQNs for simple names that map to unique FQNs or multiple FQNs? ??Cut?}

    \item RQ4 - How well does the FQN inference on CoPilot perform on real-world partial code,  compared with the state-of-the-art prompt-tuning based FQN inference method~\cite{Huang2022PrompttunedCL}?
    %s can address the shortcomings of existing code language models in FQN inference tasks.

    %\item RQ5 - Are there any interesting serendipitous findings about human-copilot collaboration design?
\end{itemize}

Next, we describe our datasets, prompt-engineering configurations, preparation of task inputs for large-scale experiments, our experiment environment and process.

\subsection{Datasets}\label{Sec:Evaluation Dataset}

%To ensure that the dataset is sufficiently diverse and consistent with the distribution of FQN predictions in practice, 
To obtain accurate and practical answers to our research questions, we construct two datasets:
one Github code dataset from the six libraries (Android SDK, JDK, GWT, Hibernate, Joda Time and Xstream) and one SO dataset from Stack Overflow posts discussing the API usage of these six libraries.

\subsubsection{Github Dataset From Six Libraries}\label{Sec:Evaluation Dataset From Github Library Source Code}
%To ensure sufficient variety in the dataset, 
We download the source code of the six libraries from the replication package provided by~\cite{Huang2022PrompttunedCL}.
The original dataset was downloaded from the library's Github repository and was used to evaluate the prompt-tuned FQN inference method in~\cite{Huang2022PrompttunedCL}.
%\zc{The six libraries have in total ??????? methods.}
%\zc{This dataset contains ??, ??, ??, ?? , ?? and ?? methods from Android SDK, JDK, GWT, Hibernate, Joda Time and Xstream, respectively.}
%The six libraries are \textit{Android}, \textit{JDK}, \textit{GWT}, \textit{Hibernate}, \textit{Joda Time} and \textit{Xstream}.

We extract all methods (not just public methods) from each source file, because all methods can be used as the FQN usage context no matter their visibility.
The body of each method is considered as a code snippet in our experiments.
As the source code of these libraries is compilable, we collect unique $<$simplename, FQN$>$ pairs used in the library methods using the Spoon~\cite{SpoonTool} tool.
These $<$simplename, FQN$>$ pairs provide the ground-truth for the large-scale experiments of the CoPilot's FQN inference capability in RQ1, RQ2 and RQ3.
A long method generally includes many $<$simplename, FQN$>$ pairs, which will generate too many experiment instances through the combination of different in-context learning factors.
Through a pilot study of manual effort, we decide to keep only the methods with less than 30 lines of code (LOC), which account for 93.42\% of all the methods in the six libraries.

% \vspace{-0.5mm}
Because CoPilot can only be invoked manually in the IDE editor, we cannot perform automatic experiments as in~\cite{Huang2022PrompttunedCL} due to the prohibitive manual effort required.
%, so we must rely on manual crowdsourcing to obtain the results of copilot tests.
%It would incur prohibitive manual effort and also be error prone to test all the methods in the six libraries in dozens of learning configuration variants. 
Therefore, we sample a subset of diverse and representative methods from the original dataset.
%and each RQ involves many variants of in-context learning configurations
%The library source code is compilable which provides the ground-truth FQNs t
First, we randomly sample a package and randomly sample a method in this package and add the method to the sample dataset.
Then, we randomly sample the methods in the other not-yet-sampled packages one at a time as follows.
We randomly sample a not-yet-sampled package and collect as candidates all methods in the package whose similarity is less than 0.9 with any of the methods in the sample dataset.
The method similarity is measured by the code embedding method~\cite{Ye2016FromWE}.
We add the least-similar candidate method with more than three $<$simplename, FQN$>$ pairs to the sample dataset, as those with too few $<$simplename, FQN$>$ pairs are not sufficient for certain learning configurations (e.g., few-shot (random example prompts).
This sampling process continues until all the library packages are
iterated.

We obtain 1,440 methods as our Github dataset, which have 8,258 $<$simplename, FQN$>$ pairs (including 3,871 unique simple names and 4,697 unique FQNs from 850 packages).
%As shown in the sampled column in Table ~\ref{Table:Diversity and Coverage of Dataset}, 
%\zc{The sampled FQNs come from 1,440 distinct library packages, accounting for 67\% of all the unique FQN packages in the original dataset. ??this needs to be clarified. why 1440 method from 1440 package? 1440 is library packages or the package names of used FQNS?}\liao{package name is the string *** in "package ***;" at the beginning of each java file. Only 1440 package names have been used because some of the methods corresponding to package names have more than 0.9 similarity to the methods in the sample dataset。
%And the number of unique FQNs in the methods corresponding to package names is less than 3, and cannot form more than 3 $<$simplename, FQN$>$ pairs.
%We only get one random method from each package name, but we try to ensure that the package name has a wide range. This is because a wide range of package names ensures the diversity of FQNs.}
We confirm that the sampled methods are representative in terms of code LOCs, FQN lengths, usage times and simplename-FQN cardinalities, and also diverse in terms of low pair-wise code similarities and FQN-set Jaccard coefficients.
Details are reported in the Section~I-A and Section~I-B in the supplementary document.

% \vspace{-1mm}
\subsubsection{Dataset From Stack Overflow Posts}\label{sec:Evaluation Dataset From Stack Overflow Source Code}
In RQ4, we would like to evaluate CoPilot's FQN inference capability in real-world partial code developers write.
To that end, we use the two partial code datasets (Stat-Type-SO and Short-SO) collected from the Stack Overflow posts discussing the API usage of the six libraries.
In our previous work~\cite{Huang2022PrompttunedCL}, we manually labelled the ground-truth FQNs for the simple names in the code snippets.
We double-check and confirm the label correctness.
The dataset Stat-Type-SO has been used in several FQN inference studies~\cite{Phan2018StatisticalLO, Saifullah2019LearningFE, Dong2022SnRCT, Huang2022PrompttunedCL}.
It contains 381 code snippets (LOCs from 3 to 30) which use 685 distinct APIs of the library APIs from 35 distinct packages.
The Short-SO was created by~\cite{Huang2022PrompttunedCL}.
It contains 115 short partial code snippets (LOC $\leq$ 10) which use 205 distinct APIs from 35 distinct packages.
In total, the two SO datasets have 2,320 $<$simplename, FQN$>$ pairs (including 684 unique simple names, 791 unique FQNs in 496 code snippets.).
%The distribution of FQN lengths, usage times and simplename-FQN cardinalities can be found in Table~\ref{tab:??rq3}.

\begin{table}[t]
    \centering
    \caption{\scriptsize Two Configurations of Prompt Engineering}
    % \vspace{-3mm}
    \resizebox{\linewidth}{!}{
    \begin{tabular}{|c|c|c|}
        \hline \textbf{Prompt Factor} & \textbf{Basic} & \textbf{Best}\\\hline
        Code Context & Provided & Provided\\\hline
        Task Description & Verbose & Concise\\\hline
        %Amount of Example Prompts & Few-Shot (REP) \\\hline
        Prompt Template & Description & Description\\\hline
        \makecell[c]{Example Prompt Order} & Random Order & Infrequent First\\\hline
        Identifier Format & With Quote & With Quote\\\hline
    \end{tabular}
    }
    \label{Table:configurations of prompt engineering}
    % \vspace{-9mm}
\end{table}

% \vspace{-1.5mm}
\subsection{Configurations of Prompt Engineering}
% \vspace{-1mm}

%We refer to the configuration of five prompt-engineering factors as shown in 
Before our experiments, we design a basic configuration (the \textit{Basic} column in Table~\ref{Table:configurations of prompt engineering}) based on our intuition of which options would be most effective for FQN inference.
The basic configuration provides the code context, uses verbose task description (i.e., ``parse simple name to fully qualified name''), provides example prompts in random order, and adopts description-style prompt template and annotates simple names and FQNs with quotes (i.e., the fully qualified name of ``simplename'' is ``FQN'').
To study the impact of one prompt design factor on the FQN inference, we vary the concerned factor with different variants but leave other factors the same as the basic configuration.
We validate our intuition and identify the best configuration in RQ1~\ref{sec:RQ1 sensitivity} (the \textit{Best} column in Table~\ref{Table:configurations of prompt engineering}).
%shows the best configuration of the five prompt engineering factors.
The best configuration is consistent with the basic configuration except for using concise task description (i.e., ``type inference'') and providing the infrequent FQN examples first.
This best configuration is used in RQ2/3/4.
In all experiments, we apply a prompt-engineering configuration in the five different shot settings.

% \vspace{-2mm}
\subsection{Composing Task Inputs for Large-Scale Experiments}
% \vspace{-1mm}

To conduct large-scale experiments, it is impractical to ask developers to manually write the task inputs in the IDE.
Therefore, given a code snippet in our datasets, we automatically compose the task inputs according to a learning configuration.
The composed task inputs are saved in .java files and stored in a folder structure corresponding to the factor options.
In this work, we assume the same simple name appearing at different places in the code context refers to the same FQN.
That is, there is no variable shadowing or name masking.
For each unique $<$simplename, FQN$>$ pair in the code context, we produce a to-be-complete prompt for the simplename with the FQN as the ground truth.
Then, we prepare the example prompts with the rest $<$simplename, FQN$>$ pairs for five different shot settings.
The composed example prompts will be ordered according to the example prompt order configuration.
The Github dataset has 8,258 $<$simplename, FQN$>$ pairs, so we compose 41,290 (5*8258) task inputs for a learning configuration. 
RQ1, RQ2 and RQ3 involve 9 configurations so we compose in total 371,610 task inputs.
The SO dataset has 2,320 $<$simplename, FQN$>$ pairs, and we compose 11,600 task inputs for the best configuration in RQ4.

% \vspace{-2mm}
\subsection{Experiment Environment and Process}\label{sec:experiment environment}
% \vspace{-1mm}

We prepare nine computers for crowd workers to manually request the completion of CoPilot.
We install JetBrain's IDEA IDE version 2022.2 on each computer and install the CoPilot plugin version 1.1.28.1744 in the IDEA environment.
As CoPilot treats source code as text, there is no need for compiling the task inputs.
We recruit nine undergraduate students from our school to complete the tasks.
We divide the task inputs roughly evenly across the nine workers.
The workers are offered a small financial incentive for their work. 

The workers open our stored task input files, obtain the predicted FQN by pressing \textit{tab} on the to-be-complete prompt line, and save the file.
We write a program to automatically extract the text after the to-be-complete prompt in the input files.
If the content is empty, the input files are missed by the workers, as CoPilot will output ``No completions were found'' if its generation fails.
We remind the workers to complete the missed inputs.
The process of completing all task inputs took 1,512 man-hours (on average 14.2 seconds per input).
%To ensure no miss of task inputs, we write a program that automatically detects whether a prompt file is successfully processed by checking if the content after the to-be-complete prompt is empty.
%This program fetches the string after the to-be-complete prompt in the prompt file and takes the first unbroken sequence in the string.
%If we get this sequence empty, we think that the prompt file was not successfully processed.
%We store all the unprocessed prompt files in a folder, and give all the files in this folder to them to generate FQNs in the second round.
%We cycle through the process of testing and reworking until there is no file in the exported folder.
Then, we extract the predicted FQNs for all model inputs.
If the content shows ``No completions were found'', the predicted FQN is marked as ``...'' which does not match any ground-truth FQNs.
If the predicted FQN contains brackets (e.g., (), [], $<>$), we keep the brackets but remove the contents inside the brackets.
If the delimiter between two tokens is a special symbol (e.g., \textit{\#}, \textit{\$}) but not the dot (\textit{.}), we replace the symbols with the dot.
As CoPilot is interactively used by developers, we believe developers can easily recognize and fix these trivial errors.
After the post-processing, we consider a predicted FQN correct only if it is identical to the ground-truth FQN.

% \vspace{-2mm}
\section{Experiment Results}

%We report our experiment results for the four RQs. 

% \vspace{-1mm}
\subsection{Sensitivity to Prompt Engineering (RQ1)}\label{sec:RQ1 sensitivity}

\begin{table*}[t]
    \centering
    \caption{Results of Sensitivity to Prompt Engineering (+/-value Against the Basic Configuration) 
    }
    % \vspace{-3mm}
    \resizebox{\linewidth}{!}{
    \resizebox{\linewidth}{!}{
    \begin{tabular}{|c|c|c|c|c|c|c|}
        \hline  PE Factor & Variant & Zero-Shot & One-Shot-ENIC & One-Shot & Few-Shot-REP & Few-Shot-LOO \\\hline
         & Basic Configuration & 49.00\% & 49.72\% & 61.18\% & 74.10\% & 77.55\% \\\hline
         
        \multirow{1}{*}{} 
        & Best Configuration & \textcolor{green!70!black}{+1.07\%} & \textcolor{green!70!black}{+0.54\%} & \textcolor{green!70!black}{+4.54\%} & \textcolor{green!70!black}{+2.30\%} & \textcolor{green!70!black}{+1.79\%} \\\hline \hline
        
        Code Context & Not Provided & \textcolor{red!90!black}{-9.00\%} & \textcolor{red!90!black}{-4.87\%} & \textcolor{red!90!black}{-4.33\%} & \textcolor{red!90!black}{-5.24\%} & \textcolor{red!90!black}{-5.03\%} \\\hline \hline
        \multirow{2}{*}{Task Description} 
        % & Verbose & 49.00\% & 49.72\% & 61.18\% & 74.10\% & 77.55\% \\\cline{2-7}
        & Concise & \textcolor{green!70!black}{+1.07\%} & \textcolor{green!70!black}{+0.54\%} & \textcolor{green!70!black}{+0.94\%} & \textcolor{green!70!black}{+1.38\%} & \textcolor{green!70!black}{+1.10\%} \\\cline{2-7}
        & No & \textcolor{red!90!black}{-1.90\%} & \textcolor{green!70!black}{+1.44\%} & \textcolor{green!70!black}{+1.34\%} & \textcolor{green!70!black}{+0.47\%} & \textcolor{green!70!black}{+0.93\%} \\\hline \hline
        
        \multirow{1}{*}{\makecell[c]{Prompt Template}} 
        % & Description & \textbf{49.00}\% & \textbf{49.72}\% & \textbf{61.18}\% & \textbf{74.10}\% & 77.55\% \\\cline{2-7}
        & Symbol & \textcolor{red!90!black}{-1.93\%} & \textcolor{red!90!black}{-0.19\%} & \textcolor{red!90!black}{-2.25\%} & \textcolor{red!90!black}{-1.47\%} & \textcolor{green!70!black}{+4.47\%}\\\hline \hline
        
        \multirow{2}{*}{\makecell[c]{Example Prompt\\Order}} 
        % & Random Order & 49.00\% & 49.72\% & 61.18\% & 74.10\% & 77.55\% \\\cline{2-7}
        & \makecell[c]{Frequent First} & - & - & \textcolor{red!90!black}{-7.18\%} & \textcolor{red!90!black}{-3.65\%} & \textcolor{red!90!black}{-1.56\%} \\\cline{2-7}
        & \makecell[c]{Infrequent First} & - & - & \textcolor{green!70!black}{+7.69\%} & \textcolor{green!70!black}{+2.95\%} & \textcolor{green!70!black}{+0.06\%} \\\hline \hline
        
        \multirow{1}{*}{Identifier Format} 
        & Without Quote & \textcolor{red!90!black}{-6.89\%} & \textcolor{red!90!black}{-2.81\%} & \textcolor{red!90!black}{-2.06\%} & \textcolor{red!90!black}{-1.06\%} & \textcolor{red!90!black}{-0.50\%} \\\hline

    \end{tabular}
    }
    }
    \label{Table:prompt engineering}
    \vspace{-5mm}
\end{table*}

% \vspace{-1mm}
\subsubsection{Motivation}

In-context learning prompts CoPilot how to complete new tasks with examples.
The model's task completion capability can be sensitive to the design of prompts (so-called prompt engineering)~\cite{liu2021pre,ding2021prompt,zhao2021calibrate}.
Our in-context learning involves five prompt-engineering factors (see Section~\ref{sec:promptengineering}), and each factor has some variants.
This RQ aims to investigate the CoPilot's sensitivity to prompt engineering and validate our intuition of basic configuration. 
%To investigate how to maximize the activation of FQN knowledge learned by Copilot while solving FQN inference tasks, we change the prompt method from four aspects: Task Description, Answer Format, Example Order, and Identifier, and observe how the accuracy of the inferred FQNs changes.
%And, we want to explore how much potential the model can play in predicting FQN with all components at the best.\liao{If we need to add a setting to test the accuracy while all the best configuration?}

\subsubsection{Methodology}
% There are three types of task descriptions: no description, concise description(e.g., FQN inference) and clear description (e.g., parse simple name to fully qualified name).
%For each component variable, we fix all components in basic configurations except itself, and generate the prompt based on five shot types for each $<$simplename, FQN$>$ pair of the dataset from Github.
We use the Github dataset in this RQ.
We create variant configurations by varying one of the five factors (code context, task description, prompt template, example prompt order or identifier format) and keeping the other four factors the same as the basic configuration.
%We have two variants for code context (with/without), three variants for task description (no/concise/verbose), two variants for prompt template (description or symbol), three variants for example prompt order (random, frequent-first or infrequent-first), and two variants for identifier format (normal word or word with quote).
In addition to the basic configuration, we obtain 7 more configurations by factor ablation, and identify the best configuration of factor settings.
We experiment each configuration with five different shot settings which produce the final 45 variant configurations, with the total 371,610 model inputs to complete.

\subsubsection{Result Analysis}

% \zc{Move code context results as part of prompt engineering. I write up some discussion. Please check.}
The results of with/without code context confirms our intuition.
Code context is critical for making context-sensitive inference.
Without code context, CoPilot essentially generates FQNs based on its memory of the correlations between simple names and FQNs~\cite{Elazar2022MeasuringCE}, and suffers the largest accuracy drop.
However, the inference accuracy without code context is still acceptable (the upper bound 72.52\% at Few-Shot-LOO).
This does indicate that CoPilot memorizes many FQNs.
On the other hand, simply relying on model memorization may bias the FQN inference by the data distribution for model pre-training.
For example, without code context, CoPilot always generates \textit{java.util.Date} for \textit{Date}, while given the SQL processing code, it will generate \textit{java.sql.Date}.
As \textit{java.util.Date} is more frequently used than \textit{java.sql.Date}, CoPilot prefers the former over the latter when no code context is provided.

The results of task description variants do not follow our intuition.
Verbose task description is only better than no task description at Zero-Shot (i.e., no example prompt provided).
For all other cases, verbose task description is worse (around 1\%) than concise and no task description.
The default example prompt template (``the fully qualified name of ... is ...'') carries similar information as verbose task description, which may reduce the importance of verbose task description.
%\zc{To confirm this hypothesis, we carry out follow-up experiments in which we combine task description variants with symbol-based prompt template (i.e., simplename $\rightarrow$ FQN and keep the other three factors the same as the basic configuration.
%In these configurations, verbose task description is better than concise/no description ... ??can we do some follow-up experiments? the results can be presented in supplementary material.}\liao{I think there is maybe no time to do this...}

The results of two prompt templates are somewhat surprising.
The description template (``the fully qualified name of ... is ...'') is better (around 2\%) than the symbol template (``simplename $\rightarrow$ FQN''), except for Few-Shot-LOO where the symbol template is 4.5\% better than the description template.
This suggests that unless there are sufficient example prompts, natural language examples are more beneficial for conditioning the model on the task.
However, when example prompts are sufficient, the model can derive the meaning of the symbol in the task context and complete the task more correctly.

The results of example prompt order are very interesting.
Frequent-first (i.e., more-frequently-used FQNs appear at the beginning) is always the worst, random-order is in the middle, and infrequent-first (less-frequently-used FQNs appear at the beginning) is always the best.
The accuracy (77.05\%) of infrequent-first at Few-Shot-REP is even better than the accuracy (75.99\%) of frequent-first at Few-Shot-LOO.
That is, letting the model see more challenging (even relatively) examples first is beneficial for the model to learn better and faster.
%However, this may not translate into practice, 
%Although developers may only know some frequently-used FQNs in a code snippet, as usage frequency is relative, it would still be beneficial to put relatively less-frequently-used FQNs before most-frequently-used FQNs.
The variants of example prompt order is not applicable to Zero-Shot (no example prompt) and One-Shot-ENIC (the same $<$simplename, FQN$>$ not in the code context).

The results of identifier format confirm our intuition.
Using a special symbol (``'' in this work) to annotate simple names and FQNs results in better inference accuracy, especially for Zero-Shot, One-Shot-ENIC and One-Shot.
The effect of special symbol diminishes as the example prompts increases.
These results suggest that special symbol let the model attend to the key information (simple name and FQN) when there is limited information.
However, when there are several example prompts, the model can distinguish the key information (different across examples) from other repeating information even the key information is not specially annotated.

Based on the prompt engineering results, we define the ``best'' configuration: with code context, concise task description, description-style prompt template, infrequent-first example prompt order, identifier with quote.
We carry out the experiments with the best configuration.
The best configuration is always better than the basic configuration across the five shot settings.
It is 1.07\%, 0.54\%, 4.54\%, 2.30\%, 1.79\% more accurate in zero-shot, one-shot-EXIC, one-shot, few-shot-REP, and few-shot-LOO, respectively.
The best configuration is also generally better than the basic configuration with only one factor variant, except for four configuration variants (no task description at One-Shot-ENIC, infrequent-first at One-Shot, infrequent-first at Few-Shot-REP, and symbol prompt template at Few-Shot-LOO).
This suggests that prompt factors have complex interactions which may cancel the effects of others.

% \vspace{1mm}
\noindent\fbox{\begin{minipage}{8.6cm} \emph{Although studies~\cite{liu2021pre,ding2021prompt,zhao2021calibrate} show PLMs are sensitive to prompt engineering, CoPilot remains overall stable in inferring FQN knowledge in face of variant prompt-engineering factors. Our intuition of the effectiveness of factor variants largely holds, except for task description and example prompt order. This indicates the necessity to combine intuition and empirical evidences in prompt engineering. Combining the best individual variants results in an overall-balanced best configuration.} \end{minipage}}

\begin{comment}
1. A concise task description can effectively prompt the model to complete the task.
2. Using natural language to answer questions with insufficient examples is more effective, and using a formalized way of answering questions with sufficient examples allows the model to achieve the best results.
3. The examples presented from hard to easy can motivate the model to perform the task better.
4. Using special symbols can help the model pay more attention to the identifier.
\end{comment}

% \vspace{-1mm}
\subsection{Impact of Amount of Example Prompts (RQ2)}\label{sec:impact of amount of example prompts}

\begin{table*}[t]
    \centering
    \caption{The FQN Inference Accuracy For FQNs with Different Data Distribution Properties (the Closer to 1, the Brighter the Color)}
    % \vspace{-3mm}
    \resizebox{\linewidth}{!}{
    \begin{tabular}{|c|c|c||c|c|c|c|c|}
        \hline   & Range & \makecell[c]{FQN Percentage(\%)} & Zero-Shot & One-Shot-ENIC & One-Shot & Few-Shot-REP & Few-Shot-LOO \\\hline
        \makecell[c]{Best Configuration} & all & 100\% & \cellcolor[RGB]{255,127,0} 50.07\% & \cellcolor[RGB]{255,128,0} 50.26\% & \cellcolor[RGB]{255,167,0} 65.72\% & \cellcolor[RGB]{255,194,0} 76.40\% & \cellcolor[RGB]{255,202,0} 79.34\% \\\hline \hline
        
        \multirow{4}{*}{\makecell[c]{FQN Length}}
        & $2-4$ & 58.04\% & \cellcolor[RGB]{255,195,0} 76.59\% & \cellcolor[RGB]{255,197,0} 77.49\% & \cellcolor[RGB]{255,199,0} 78.40\% & \cellcolor[RGB]{255,220,0} 86.42\% & \cellcolor[RGB]{255,225,0} 88.29\% \\\cline{2-8}
        & $5-7$ & 28.00\% & \cellcolor[RGB]{255,47,0} 18.59\% & \cellcolor[RGB]{255,46,0} 18.19\% & \cellcolor[RGB]{255,129,0} 50.73\% & \cellcolor[RGB]{255,163,0} 64.25\% & \cellcolor[RGB]{255,175,0} 68.92\% \\\cline{2-8}
        & $8-10$ & 13.35\% & \cellcolor[RGB]{255,7,0} 3.08\% & \cellcolor[RGB]{255,3,0} 1.40\% & \cellcolor[RGB]{255,110,0} 43.19\% & \cellcolor[RGB]{255,151,0} 59.33\% & \cellcolor[RGB]{255,161,0} 63.25\% \\\cline{2-8}
        & $\geq 11$ & 0.61\% & \cellcolor[RGB]{255,0,0} 0.00\% & \cellcolor[RGB]{255,0,0} 0.00\% & \cellcolor[RGB]{255,104,0} 40.82\% & \cellcolor[RGB]{255,140,0} 55.10\% & \cellcolor[RGB]{255,150,0} 59.18\% \\ \hline \hline
        % & All & 100\% & 50.07\% & 50.26\% & 65.72\% & 76.40\% & 79.34\% \\\hline

        \multirow{4}{*}{\makecell[c]{FQN Usage Time}}
        & $\geq 10k$ & 12.58\% & \cellcolor[RGB]{255,253,0} 99.42\% & \cellcolor[RGB]{255,254,0} 99.71\% & \cellcolor[RGB]{255,253,0} 99.52\% & \cellcolor[RGB]{255,253,0} 99.42\% & \cellcolor[RGB]{255,254,0} 99.62\% \\\cline{2-8}
        & $[1k,10k)$ & 14.99\% & \cellcolor[RGB]{255,203,0} 79.75\% & \cellcolor[RGB]{255,219,0} 85.98\% & \cellcolor[RGB]{255,211,0} 82.99\% & \cellcolor[RGB]{255,228,0} 89.71\% & \cellcolor[RGB]{255,235,0} 92.53\% \\\cline{2-8}
        & $[10,1k)$ & 42.20\% & \cellcolor[RGB]{255,132,0} 51.87\% & \cellcolor[RGB]{255,128,0} 50.56\% & \cellcolor[RGB]{255,156,0} 61.48\% & \cellcolor[RGB]{255,192,0} 75.56\% & \cellcolor[RGB]{255,201,0} 79.18\% \\\cline{2-8}
        & $[1,10)$ & 30.23\% & \cellcolor[RGB]{255,29,0} 11.51\% & \cellcolor[RGB]{255,27,0} 10.72\% & \cellcolor[RGB]{255,123,0} 48.49\% & \cellcolor[RGB]{255,155,0} 61.04\% & \cellcolor[RGB]{255,163,0} 64.27\% \\\hline \hline
                
        % & All & 100\% & 50.07\% & 50.26\% & 65.72\% & 76.40\% & 79.34\% \\\hline

        \multirow{4}{*}{SN:FQN}
        & $1:1$ & 74.80\% & \cellcolor[RGB]{255,136,0} 53.54\% & \cellcolor[RGB]{255,138,0} 54.45\% & \cellcolor[RGB]{255,179,0} 70.48\% & \cellcolor[RGB]{255,203,0} 79.64\% & \cellcolor[RGB]{255,210,0} 82.52\% \\\cline{2-8}
        & $1:2$ & 9.53\% & \cellcolor[RGB]{255,115,0} 45.23\% & \cellcolor[RGB]{255,120,0} 47.19\% & \cellcolor[RGB]{255,153,0} 60.00\% & \cellcolor[RGB]{255,190,0} 74.64\% & \cellcolor[RGB]{255,194,0} 76.47\% \\\cline{2-8}
        & $1:3$ & 3.88\% & \cellcolor[RGB]{255,92,0} 36.22\% & \cellcolor[RGB]{255,87,0} 34.29\% & \cellcolor[RGB]{255,132,0} 51.92\% & \cellcolor[RGB]{255,174,0} 68.27\% & \cellcolor[RGB]{255,187,0} 73.40\% \\\cline{2-8}
        & $1:\geq 4$ & 11.79\% & \cellcolor[RGB]{255,93,0} 36.53\% & \cellcolor[RGB]{255,79,0} 31.36\% & \cellcolor[RGB]{255,113,0} 44.67\% & \cellcolor[RGB]{255,152,0} 59.97\% & \cellcolor[RGB]{255,161,0} 63.46\% \\\hline \hline
        % & All & 100\% & 50.07\% & 50.26\% & 65.72\% & 76.40\% & 79.34\% \\\hline

        \multirow{4}{*}{FQN:SN}
        & $1:1$ & 70.36\% & \cellcolor[RGB]{255,139,0} 54.59\% & \cellcolor[RGB]{255,141,0} 55.39\% & \cellcolor[RGB]{255,174,0} 68.57\% & \cellcolor[RGB]{255,194,0} 76.27\% & \cellcolor[RGB]{255,200,0} 78.48\% \\\cline{2-8}
        & $1:2$ & 12.95\% & \cellcolor[RGB]{255,68,0} 26.73\% & \cellcolor[RGB]{255,64,0} 25.38\% & \cellcolor[RGB]{255,146,0} 57.60\% & \cellcolor[RGB]{255,193,0} 75.87\% & \cellcolor[RGB]{255,206,0} 81.15\% \\\cline{2-8}
        & $1:3$ & 3.91\% & \cellcolor[RGB]{255,71,0} 28.03\% & \cellcolor[RGB]{255,64,0} 25.16\% & \cellcolor[RGB]{255,144,0} 56.69\% & \cellcolor[RGB]{255,196,0} 77.07\% & \cellcolor[RGB]{255,203,0} 79.94\% \\\cline{2-8}
        & $1:\geq 4$ & 12.78\% & \cellcolor[RGB]{255,141,0} 55.56\% & \cellcolor[RGB]{255,139,0} 54.87\% & \cellcolor[RGB]{255,155,0} 61.01\% & \cellcolor[RGB]{255,197,0} 77.49\% & \cellcolor[RGB]{255,209,0} 82.07\% \\\hline

    \end{tabular}
    }
    \label{Table:accuracy of different FQN data distribution properties}
    \vspace{-5mm}
\end{table*}

% \vspace{-1mm}
\subsubsection{Motivation}
In-context learning relies on example prompts to adapt the model to the new tasks unseen during pre-training.
From the practical point of view, example prompts correspond to priori FQN knowledge developers have which may help the model complete the FQN inference.
This RQ aims to investigate the CoPilot's FQN inference capability when the developer can provide different amount of priori-known FQNs.
The results help us evaluate the practicality of CoPilot for FQN inference, and estimate the extent of FQN knowledge stored in CoPilot.

\subsubsection{Methodology}\label{Sec:RQ2 Methodology}
We use the Github dataset in this RQ.
We use the best configuration of prompt factors (see Section~\ref{sec:RQ1 sensitivity}) and experiment five different amount of example prompts (see Section~\ref{sec:amount of example prompts}).
%and keep the other five factors the same as those in the best configuration of in-context learning (see Section~\ref{sec:RQ1 sensitivity}).
For each $<$$simplename$, $FQN$$>$ pair in a code snippet, we generate five model inputs, corresponding to the five shot settings respectively.
We obtain 41,290 model inputs (8,258 for each shot setting) for this RQ.

\subsubsection{Result Analysis}
The \textit{Best Configuration} in Table~\ref{Table:accuracy of different FQN data distribution properties} shows our results.
At Zero-Shot, the inference performs poorly (50.07\% accuracy).
\begin{comment}
At zero-shot, except for very short FQNs (2 to 4 words), the inference fails miserably (the accuracy below 16.94\% for the FQNs with 5 or more words).
Similarly, except for very frequently used FQNs (?? to ?? times), the inference fails miserably (the accuracy below ??\% for the FQNs used ?? or less times)
\end{comment}
A common mistake CoPilot makes at zero-shot is to generate some FQN-irrelevant text.
For example, it generates a path name ``Data.Exported\_Datas.BasicConfig-zero-shot.prompt\_files.Ticket'' for the simple name ``Ticket'', which is the file that stores the task input.
%\zc{The other common mistake is formatting errors. For example, ...``.'' as ''\$'' ... because Copilot mixes the delimiters of different naming convention ...}
%\liao{I mentioned that we will automatically replace the special symbols with `.' in section~\ref{sec:experiment environment}, so if the predicted FQN is all right except for the special symbols, we will assume it is correct.
%So I'm not sure whether I need to describe this example here.}
The poor zero-shot accuracy suggests that only a task description is insufficient for the model to determine what task it needs to solve.

One-shot-ENIC (example not in code) shows the model one example prompt.
Although this example is not for the simple name in the code context, it helps to reduce the cases where the model generates FQN-irrelevant text.
However, the overall accuracy (50.26\%) of One-Shot-ENIC does not improve much over that of Zero-Shot, because the model often generates some FQNs irrelevant to the code context at One-Shot-ENIC, which could be misled by the non-in-the-code example.
% \zc{For example, for the simple name ``acquire()'', the ground-truth FQN is ``PowerManager.WakeLock.acquire()'', but the model generates the FQN ``android.os.PowerManager\$WakeLock.acquire()'' ??this is a wrong example. The ground-truth seems wrong as the correct FQN should be ``android.os.PowerManager.WakeLock.acquire()''?????}.

Providing an example prompt for the simple name and FQN in the code context helps the model correct many irrelevant-to-context FQN inference errors, which leads to significant accuracy improvement (overall from 50.26\% at One-Shot-ENIC to 65.72\% at One-Shot).
%This improvement comes from mainly the improvements for longer or less frequently used FQNs.
Further increasing example prompts can further boost the inference accuracy (overall 76.40\% Few-Shot-REP and 79.34\% Few-Shot-LOO).
At Few-Shot-LOO, the model gives the upper bound of the model's inference capability: it correctly inferences 3,224 distinct FQNs.
% Providing more example prompts is especially helpful for inferring longer or less-frequently-used FQNs.
% At Few-Shot-LOO, the model achieves the inference accuracy 59.18\% for the FQNs with 11 or more words and the accuracy ??.??\% for the FQN used ?? or less times.

\noindent\fbox{
\begin{minipage}{8.5cm} \emph{CoPilot stores rich FQN knowledge. Its FQN knowledge can be reasonably recalled even when only one simplename-FQN example in the code context is provided.
As the number of examples increases, more and more FQN knowledge can be accurately recalled, with the maximum accuracy at about 80\%.
Therefore, it is practical to use CoPilot with in-context learning for FQN inference.} \end{minipage}}

%The more examples the model sees, the better it recalls the factual knowledge gained during pretraining, and the more it understands the situation at hand, the correct solution structure and the scope of the answer.
%The more examples the model sees, the better it can handle the task and the more it can accomplish more difficult tasks.

% \vspace{-1.5mm}
\subsection{Impact of FQN Data Distribution Properties (RQ3)}
\label{sec:RQ3}
% \zc{Another thought: putting 1:1 1:N in RQ2 will make RQ1 too long and too complex. I am now thinking still having RQ2 for shot settings only and RQ3 for data distribution properties.}

% \zc{Do you consider only Simple Type Name:Type FQN or also simple varname:FQN?}\liao{I just consider simple type name:Type FQN now.}
% \zc{Just a thought: For simple varname:FQN, we have N:1 (i.e., multiple varnames map to the same FQN). What is the percentage of simple varname:FQN N:1? Anything interesting to discucss?}

% \vspace{-1mm}
\subsubsection{Motivation}

FQN data exhibits Zipfian distribution and dynamic meanings which have been reported to be influential in the in-context learning ability of PLMs~\cite{Chan2022DataDP}.
This RQ aims to reveal the correlations between the FQN properties, different shot settings and the CoPilot's FQN inference ability.
The results help us understand the characteristics of FQN knowledge stored in CoPilot, and the conditions to effectively retrieve FQNs with different properties.

\subsubsection{Methodology}
We consider four FQN properties: lengths, usage times, and simplename-FQN (SN:FQN) and FQN-simplename (FQN:SN) cardinalities (i.e., polysemy and synonymy ambiguity respectively).
We obtain the statistics of these FQN properties in the original Github dataset (see Appendix Section~I in our supplementary document). 
%??all statistics are not from the sampled methods, right?
%\liao{The statistics in the introduction are from the original Github, while the FQN percentages in the Table~\ref{Table:accuracy of different FQN data distribution properties} result are from the sampled dataset.}
The experiment setting is the same as RQ2.
We calculate the accuracy for four FQN length ranges (2-4, 5-7, 8-10 and $\geq$11), four FQN usage time ranges [1,10), [10-1k), [1k-10k) and $\geq$10k), four SN:FQN cardinalities (1:1, 1:2, 1:3, 1:$\geq$4), and four FQN:SN cardinalities (1:1, 1:2, 1:3, 1:$\geq$4).
SN:FQN 1:1 can be different from FQN:SN 1:1 as they are indexed by unique simple names and unique FQNs respectively.
\textit{FQN percentage} column in Table~\ref{Table:accuracy of different FQN data distribution properties} shows the percentage of each range in our sample dataset.

% \zc{Put this in the supplementary materials.
% Note that SN:FQN means the number of FQNs corresponding to the same simple name, and FQN:SN means the number of simple names corresponding to the same FQN.
% The two cases are different, such as the case where SN:FQN is 1:1 is not equivalent to the case where FQN:SN is 1:1.
% If the simple name \textit{br} corresponds only to \textit{java.io.BufferedRead}, and the simple name \textit{Buffread} corresponds only to \textit{java.io.BufferedRead}, they are both 1:1 in the case of SN:FQN.
% However, in the case of FQN:SN, they are both 2:1.}

% \zc{four SN:FQN cardinalities (1:1, 1:2, 1:3, 1:$\geq$4, 2:1, 3:1 , $\geq$4:1) ??update after adding SN:FQN N:1}.
% \zc{The SN:FQN cardinality reflects the name ambiguity.}

%\vspace{-1mm}
\subsubsection{Results Analysis}
%The \textit{FQN Length} in shows the accuracy of Copilot inference for different length of FQNs.
Table~\ref{Table:accuracy of different FQN data distribution properties} shows the results.
Looking at the inference accuracy at different FQN length ranges, it is unsurprising that the accuracy degrades as the FQNs become longer.
For short FQNs (2-4 tokens), the model can make fairly accurate inference even without any example prompts (76.59\% at Zero-Shot), and achieves the maximum accuracy 88.29\% at Few-Shot-LOO.
As the FQNs become longer (5 or more tokens), the inference accuracy drops over 58\% to below 18.59\% at Zero-Shot and below 18.19\% at One-Shot-ENIC.
For FQNs with 8 or more tokens, the accuracy is close to 0\%.
%The decrease in accuracy is starting to slow as at least one example provided that the simple name and corresponding FQN in the code context.
However, with only one example in the code context, the accuracy dramatically jumps to 50.73\% for FQN length 5-7 and over 40\% for FQN length $\geq$8.
Increasing the number of examples can further boost the accuracy.
At Few-Shot-REP and Few-Shot-LOO, the accuracy for the most challenging FQN length $\geq$11 is 55.10\% and 59.18\%, respectively.

%The model cannot make (relatively) accurate inferences for longer FQNs unless some example prompts are provided at Few-Shot-REP and Few-Shot-LOO. 
%Similar observation can be made for different FQN usage frequency ranges.

For the most frequently-used FQNs ($\geq$10k), CoPilot achieves almost perfect accuracy ($\geq$99\%) even at Zero-Shot.
For the FQN usage times [1k, 10k), it also performs very well (79.75\% at Zero-Shot and 90\% or above at Few-Shot).
For less frequently-used FQNs ([10, 1k), CoPilot still maintains a reasonable inference accuracy (about 50\%-60\% at Zero-Shot and One-Shot), unlike the close-to-0 accuracy for long FQNs ($\geq$8 tokens).
For the FQNs in the ([10, 1k) range, providing some FQN examples improves the accuracy to 75.56\% at Few-Shot-REP and maximizes the accuracy to 79.18\% at Few-Shot-LOO.
For the lest frequently-used FQNs ($<$10), CoPilot has to see some FQN examples to make the inference with reasonable accuracy (above 61\% at Few-Shot).

The overall trend of accuracy changes for SN:FQN (1:N) across the five shot settings is similar to that for FQN lengths and usage times.
One difference is that CoPilot still has a certain level of inference capability for the challenging SN:FQN (1:$\geq$3) (31\%-36\% accuracy at Zero-Shot and One-Shot-ENIC), rather than the catastrophic failures for FQN length $\geq$8 and FQN usage times $<$10.
The other difference is that the accuracy gaps between the easy and the more challenging SN:FQN cases are smaller than those between different FQN length ranges and FQN usage-time ranges (visible from the color differences across the ranges).
An interesting result is that CoPilot may make mistakes for SN:FQN (1:1).
For example, the simple name \textit{Cookies}, CoPilot sometimes predicts the FQN \textit{com.google.gwt.http.client.Cookies} while the ground-truth is \textit{com.google.gwt.\underline{user}.client.Cookies}.
%In fact, the words \textit{client} and \textit{Cookies} have a high probability of being related to \textit{http}.
%, so CoPilot does not generate \textit{user} but generate \textit{http}.
We attribute this to the probabilistic nature of the neural network.
Note that even for the dictionary-lookup in a symbolic FQN knowledge base~\cite{SnRConstraint, Phan2018StatisticalLO, Saifullah2019LearningFE}, the accuracy at SN:FQN (1:1) is not 100\% either, because the symbolic knowledge base suffers from the OOV issue limited by code compilation~\cite{Huang2022PrompttunedCL}.
In contrast, CoPilot does not require any code compilation or analysis which is much more convenient to deploy in practice.

The trend of increasing accuracy from Zero-Shot to Few-Shot-LOO for FQN:SN is the same as that for the other three properties.
However, at a particular shot setting (One-Shot, Few-Shot-REP or Few-Shot-LOO), the accuracy at different FQN:SN ranges has much smaller differences, compared with the accuracy differences between different ranges of the other three properties.
That is, the number of different variable names referring to the same FQN (i.e., synonymy ambiguity) do not affect much the inference of this FQN once one FQN example in the code context is provided.
This is because the variables are used to invoke the methods or access the fields (e.g., \textit{br.readLine()}) and the method/field name provides good usage context for inferring the type of the variable name.
Another interesting difference from the other three properties is that the middle FQN:SN ranges (1:2 and 1:3) are much worse than FQN:SN 1:1 and 1:$\geq$4 at Zero-Shot and One-Shot-ENIC, and the accuracy at FQN:SN 1:$\geq$4 is in par with that at FQN 1:1 for all shot settings.
FQN:SN 1:$\geq$4 actually means more usage of an FQN although the FQN is referred by many different variable names.
The benefit of the higher usage times outweighs the challenge incurred by different variable names.

% \vspace{1mm}

\noindent\fbox{
\begin{minipage}{8.6cm} \emph{The FQN knowledge in CoPilot is diverse in FQN lengths, usage times, SN:FQN and FQN:SN cardinalities. CoPilot can accurately infers short, frequently-used, less ambiguous FQNs, without the need of many task demonstrations. Providing more task demonstrations helps CoPilot better infer longer, less frequently-used, or more ambiguous FQNs. FQN usage time is the most influential on inference accuracy, followed by FQN length, and then name ambiguity. Synonymy (FQN:SN 1:$\geq$2) is the least influential as it indicates higher FQN usage which is beneficial for FQN inference.} \end{minipage}}

% \vspace{-2mm}
\subsection{Inference for Real-World Partial Code (RQ4)}\label{sec:RQ4 compare with mlm}
% \vspace{-1mm}

%  

\begin{table*}[t]
    \centering
    \caption{\scriptsize Comparison between PCM-based FQN Inferences}

    \resizebox{\linewidth}{!}{
    \begin{tabular}{|l|c|c|c|c|}
        \hline  Method & \makecell[c]{Test Strategy} & Stat-Type-SO & Short-SO & Overall  \\\hline
        
        \multirow{3}{*}{\makecell[l]{Pre-trained CodeBERT MLM}}
        & Individuals & 18.86\% & 10.89\% & 18.16\% \\\cline{2-5}
        & \makecell[l]{Majority Win} & 14.73\% & 10.75\% & 14.07\% \\\cline{2-5}
        & \makecell[l]{Any-correct} & 18.20\% & 12.28\% & 17.22\% \\\hline
        \multirow{3}{*}{\makecell[l]{Prompt-tuned CodeBERT MLM}}
        & Individuals & 88.25\% & 80.54\% & 87.56\% \\\cline{2-5}
        & \makecell[l]{Majority Win} & 88.95\% & 82.90\% & 87.94\% \\\cline{2-5}
        & \makecell[l]{Any-correct} & \textbf{89.76\%} & \textbf{82.90}\% & \textbf{88.61}\% \\\hline
        
        \multirow{5}{*}{\makecell[l]{\makecell[l]{Copilot with In-Context Learning}}}
        & \makecell[c]{Zero-Shot} & 76.09\% & 78.76\% & 76.98\% \\\cline{2-5}
        & \makecell[c]{One-Shot-ENIC} & 73.82\% & 79.35\% & 75.73\% \\\cline{2-5}
        & \makecell[c]{One-Shot} & 83.89\% & 88.79\% & 84.70\% \\\cline{2-5}
        & \makecell[c]{Few-Shot-REP} & 88.66\% & \textbf{92.04}\% & 88.88\% \\\cline{2-5}
        & \makecell[c]{Few-Shot-LOO} & \textbf{89.01}\% & 91.15\% & \textbf{89.31}\% \\\hline

    \end{tabular}
    }
    \label{Table:Comparing with MLM for FQN inference in accuracy}
    
\end{table*}

\subsubsection{Motivation}

RQ2 and RQ3 use the methods in library source code as code snippets.
As the library source code is compilable, we can automatically collect the ground-truth FQNs for the large-scale experiments in RQ2 and RQ3.
Although CoPilot performs well on library methods, we want to further confirm its FQN inference capability for real-world partial code.
Furthermore, we want to compare CoPilot's capability with that of small-size PCMs and the capability of in-context learning versus supervised prompt tuning.

%Our method helps the model to adapt quickly and efficiently to new tasks without updating weights (fine-tuning).
%To verify whether our method can outperform existing methods without fine-tuning, we compare the accuracy with an existing MLM(pre-trained CodeBERT).
%To verify whether our method can be more effective than the fine-tuning approach, we compare it with the existing MLM fine-tuning-based approach.

\subsubsection{Methodology}
We use the two Stack Overflow datasets Stat-Type-SO and Short-SO (see Section~\ref{sec:Evaluation Dataset From Stack Overflow Source Code}).
%which contain partial code snippets collected from Stack Overflow posts.
We use the best configuration identified in RQ1 (see Table~\ref{Table:configurations of prompt engineering}).
The two SO datasets contain 496 partial code snippets and 2,320 $<$simplename, FQN$>$ pairs, and we obtain 11,600 task inputs.
%We generate one model input for each pair in the five different shot settings (11,600 model inputs in total).
%We enter the model inputs in the IDE and request CoPilot to generate the FQNs.

We consider two baselines:
% First, pre-trained CodeT5~\cite{wang2021codet5} with in-context learning at five different shot settings.
% CodeT5 is a text generation model and has 220M parameters.
% We feed the same task inputs to the frozen CodeT5 and request it to generate the FQNs in the same way as the in-context learning on CoPilot.
% We do not fine-tune CodeT5 because it itself is a complex work as shown in~\cite{Huang2022PrompttunedCL}.
1) pre-trained CodeBERT~\cite{Feng2020CodeBERTAP} without fine-tuning (i.e., zero-shot);
2) pre-trained CodeBERT with FQN prompt fine-tuning as~\cite{Huang2022PrompttunedCL}.
We use the best FQN-prompt-tuned CodeBERT model in~\cite{Huang2022PrompttunedCL} (tuned with 11,776 source code files of the six libraries).
CodeBERT is a MLM and has 125M parameters.
We formulate the FQN inference on CodeBERT as a text fill-in-blank task as in~\cite{Huang2022PrompttunedCL}.

%to test the accuracy of our method, pre-trained CodeBERT~\cite{Feng2020CodeBERTAP} and the existing fine-tuned model method~\cite{Huang2022PrompttunedCL} for FQN inference.
%We implement the two models to generate FQNs for simple names.

%For all three methods, the generated FQNs and the ground-truth FQNs are matched to calculate the inference accuracy.
As illustrated in Fig.~\ref{fig:approach figure}, CoPilot makes one inference for each unique simple name in the code context.
However, CodeBERT-based methods make one inference for each individual simple name in the code context.
It may infer different FQNs for the same simple name at different locations.
We consider three accuracy variants for CodeBERT-based methods (individual instance, majority-win or any-correct).
Individual instance is the accuracy for all individual simple names.
Majority-win calculates the accuracy based on the majority of inferred FQNs for each unique simple name.
If there is a tie, an FQN is randomly picked.
Any-correct means if any of the inferred FQNs for a unique simple name is correct, we consider the model makes the correct inference.

\subsubsection{Result Analysis}

% CodeT5 with in-context learning performs the worst. 
% It does not make any correct FQN inferences, no matter which shot settings.
% CodeBert without fine-tuning performs somewhat better than CodeT5 with in-context learning, but its accuracy is still very low (20\%).
CodeBERT without fine-tuning performs the worst, with an accuracy of less than 19\% at best.
This suggests that small-size PCMs do not capture much FQN knowledge as giant CoPilot.
% , and in-context learning alone cannot adapt a PCM to make accurate FQN inference.
In contrast, standing on the shoulder of giant CoPilot which stores rich FQN knowledge, in-context learning achieves above 76\% accuracy at Zero-Shot on the two SO datasets.
Furthermore, the accuracy on the SO partial code at Zero-Shot is much higher than that on library methods at Zero-Shot (about 50\%).
This is because the FQNs in the partial code on Stack Overflow are mostly frequently-used APIs, which are easy to make accurate inference as shown in RQ3.

The FQN prompt-tuning significantly boosts CodeBERT's inference accuracy to above 80\%.
We see small fluctuations between the three accuracy variants (individual-instance, majority-win and any-correct), which suggests that CodeBERT with prompt tuning generally makes consistent FQN inferences for the same simple name at different locations.
Considering SO partial code uses many commonly used APIs, it is reasonable to assume developers would know some of the used APIs.
Therefore, One-Shot and Few-Shot-REP will reflect CoPilot's capability in practice.
At these two shot settings, CoPilot achieves 8\% and 12\% higher accuracy than Zero-Shot, and the accuracy is close to or better than that of CodeBERT with prompt tuning.
Maximizing priori-known FQNs at Few-Shot-LOO only slightly improves the accuracy.
We analyze the accuracy for the code of six libraries individually, and find that in-context learning is more stable than the prompt-tuning method.
Due to the space limitation, we put the result details in Section~II in the supplementary materials.

% \vspace{-0.3mm}
\noindent\fbox{
\begin{minipage}{8.5cm} \emph{CoPilot exhibits emergent FQN inference ability which does not exist in small PCMs. Reusing online code snippets is a common practice~\cite{rahman2018evaluating, gopstein2018prevalence, yin2018learning}. CoPilot can facilitate this code reuse by accurately infer FQNs for cannot-be-resoled simple names in uncompilable partial code.
It achieves the accuracy with only a few examples of the FQN inference task, in par with CodeBERT with FQN-prompt tuning~\cite{Huang2022PrompttunedCL}
} \end{minipage}}

%% file: Dis.tex
% \vspace{-0.5mm}
\section{Discussions}
\label{sec:discussion}
% \vspace{-2mm}

%In exploring the interaction of Copilot and humans, we get four serendipitous findings.
We now discuss our exploratory interaction experiences with CoPilot, the differences between fine-tuning and in-context learning, and potential threats to the validity of our study.

% \vspace{-2.3mm}
\subsection{Human-CoPilot Interaction}
\label{sec:humancopilotinteraction}
% \vspace{-1.5mm}

During our experiments, we perform some exploratory interactions with CoPilot inspired by its outputs.
The results of such exploratory interactions are not counted in our RQs, but they inspire ways to help CoPilot better serve the SE tasks and call for further research in human-CoPilot interaction.

% \vspace{-0.5mm}
\subsubsection{Micro-Level Sensitivity}

%(e.g., the fully qualified name of ``VMID'')  (e.g., the fully qualified name of xxx is:)
Although CoPilot remains stable in face of prompt variants (RQ1), it is sometimes sensitive to micro-level input changes.
For example, when CoPilot returns ``no completion were found'', we may trigger a successful completion by slightly changing the to-be-complete prompt, for example, delete the ending ``is'' , append a ``:'', or append several spaces.
Furthermore, CoPilot sometimes generates an FQN followed by a code snippet. 
We find that inserting an empty line between code context and task description often forces CoPilot to generate only FQNs. 
Such micro-level sensitivities of CoPilot seem inevitable, but knowing them will enhance pragmatic use of CoPilot in the SE tasks.

\subsubsection{Error Correction as Task Demonstration}

In our experiments, we complete the inference tasks without human intervention.
%determine the inference accuracy by mechanically matching the generated FQNs and the ground-truth FQNs.
However, We find that many of CoPilot's errors can be easily recognized and fixed by human.
%For example, Copilot may generate ``\$'' as separator instead of Java's ``.''.
For example, CoPilot generates only a partial FQN (e.g., \textit{org.apache.xalan.xsltc.compiler} for the simple name \textit{ClassGenerator}).
When pressing tab once more to request a further completion, CoPilot generates the full correct FQN.
%\textit{org.apache.xalan.xsltc.compiler.util.ClassGenerator}.
As another example, CoPilot sometimes generates some noisy characters in the FQNs, especially when the simple name has symbols ([], $<>$ or ()).
For example, it generates \textit{[Ljava.lang.String;} for \textit{String[]}.
The outputs of CoPilot can never be perfect, and we may not need it to be perfect~\cite{Weisz2021PerfectionNR}.
But we may teach CoPilot to avoid common errors by human feedback.
For example, human may tell CoPilot the generation is incomplete or remove the noise characters as further task demonstrations.
%Human can recognize the noise and remove the leading ``L'' and the ending ``;'' and add ``[]'' at the end.
%For example, after Copilot generates a correct FQN, developers can approve it by entering a comment ``FQN is Correct!''.
%When it generates some errors, developers can point out the errors and demonstrates the corrections, for example, ``\$ is wrong. Use . instead.''.
%Human feedback may also encourage good outputs.
%For example, Copilot occasionally generates not only the correct FQN but also what is not correct, such as ``\textit{java.io.File, not android.os.File} for  \textit{File}.
%We may encourage this useful behavior to help developers distinguish types with the same simple type name.

% \vspace{-3mm}
\subsubsection{Various Forms of Priori Knowledge}
\vspace{-0.5mm}

In our experiments, a task input includes fixed priori-known FQNs and only one simple name for inference.
%Furthermore, we generate one FQN for one simple name at a time, independent of other simple names in the code context.
In practice, developers can iteratively infer FQNs for multiple simple names and provide feedback during the process, for example by confirming the correctly-inferred FQNs or correcting the wrongly-inferred FQNs, which may help subsequent inferences.
%The correctly-generated FQNs (confirmed by developers) or the incorrect FQNs but corrected by developers in previous requests could be used as priori-known FQNs to help the inference in subsequent requests.
Other types of priori knowledge could also be useful.
For example, the developer may know the partial package name of the APIs.
Even providing just some beginning words such as \textit{com.android}, it not only shortens the FQN generation, but also conditions CoPilot on what to generate.
Or the developer may forget some parts of an FQN, she may specify the forgotten parts of the FQN by masks (e.g., ``\_'').
% ``complete the FQN \textit{com.android.\_.\_.\zc{givearealclassname ??this should be a real-world class name}}'' where Copilot can fill in the black \_.
For example, CoPilot generates \textit{com.android.\underline{layoutlib}.\underline{bridge}.\underline{impl}.\underline{binding}.AdapterItem} based on the to-be-complete-prompt ``complete the FQN \textit{com.android.\_.\_.\_.\_.AdapterItem} : ''.

%Another useful priori knowledge is the FQN length.
%Developers may experiment the typical FQN lengths of a library.
%They may indicate the length by masks in the prompt, for example, ``Complete the fully qualified name mask.mask.String'', where mask could be some special symbol (e.g., \_).
%This will condition Copilot to generate FQNs of specified length.
%, which could be priori-known as the FQN lengths of a library is limited.

% \begin{table*}[t]
%     \centering
%     \caption{In-context Learning vs. Supervised Fine-Tuning 
%     }
%     \begin{tabular}{|c|c|c|}
%     \hline & In-context Learning (ours) & Supervised Fine-Tuning (Ref{[}35{]})     \\ \hline
%     Model & CoPilot (175B) & CodeBERT (120M) \\ \hline
%     Data Collection & Not Applicable & Library Source Code \\ \hline
%     Gradient Update & NO & YES \\ \hline
%     Prompt Design & YES & NO \\ \hline
%     Extension (more Libs/PLs) & Prompt Reuse & More Data Collection and Gradient Update \\ \hline
% \end{tabular}
%     \label{table: In-context learning vs fine-tuning}
% \end{table*}

\subsection{In-context Learning vs. Supervised Fine-Tuning}
% \subsubsection{Drawbacks of Supervised Fine-Tuning Method}
% In our experiment result of RQ4 (Section ~\ref{sec:RQ4 compare with mlm}), our in-context learning method achieves the accuracy with only a few examples of the FQN inference task, in par with CodeBERT with FQN-prompt tuning~\cite{Huang2022PrompttunedCL}.
% Our key contribution is not to compete with fine-tuned models in inference accuracy, but to design a new lightweight in-context learning on frozen giant LM to achieve the same objective as model fine-tuning~\cite{Huang2022PrompttunedCL}.

The supervised fine-tuning method has some drawbacks when compared to the lightweight and easily applicable in-context learning method.

First, fine-tuning typically necessitates the collection and processing of datasets.
Our previous work~\cite{Huang2022PrompttunedCL}, for example, collect the library's source code, remove the noise in codes, and then build the dataset for tuning the pre-trained CodeBERT.
In contrast, the in-context learning method only needs a few task demonstrations to teach the frozen giant code model, and does not need to go through the data process.

Second, fine-tuning necessitates gradient update, which usually requires testing a variety of different hyper-parameters, resulting in a significant amount of work. In contrast, the in-context learning method directly activates the frozen giant code model without a gradient update.

Third, the fine-tuned model is applicable to a domain-specific task. Assume that developers discover a new and distinct downstream task.
In that case, they must repeat the model tuning processes to obtain a new domain-specific model.
In contrast, if a new scenario arises, the in-context learning method only requires reusing the prompt design to construct the demonstrations for the new task.
% Fourth, the fine-tuning methods are usually based on small models such as CodeBERT (125M) and CodeT5 (220M).
% Differently, in-context learning can only show better results based on giant models.
% This is because model size matters and emergent abilities will not appear until a critical threshold of scale is reached.

So, instead of gathering new data and fine-tuning the model, the SE task shifts to designing/reusing a prompt for extracting knowledge from frozen pre-trained LMs, which is an entirely new paradigm for leveraging the superpower of giant models for previously unseen tasks.

\subsection{Threats to Validity}
% We experiment only six libraries of Java.
% Although Java is a popular and influential language, and the six libraries include two popular large SDKs and four small-to-large libraries for diverse functionalities, further studies on other libraries and languages are required to generalize our findings.
Due to the prohibitive manual effort required, we sampled the datasets from the six libraries of Java.
It is impossible to achieve the exact same distribution as the original code, but we made our best effort to ensure the distributions of sampled code are as close as possible to those of the original code and share the same overall trend.
To ensure the diversity of the sampled dataset, we chose the 0.9 threshold for removing clone codes.
We experimented the threshold from 0.8-0.95 and found 0.9 produces the code samples with the most diverse distributions (e.g., code LOCs, FQN lengths, usage times and simplename-FQN cardinalities) which are close to the original dataset.

Due to the prohibitive manual effort, we experiment 9 prompt configurations through factor ablation, among which we identify a ``best'' configuration.
We will experiment more configurations (e.g., verbose task description with symbol-style prompt) in the future.
Our experiments show the practicality of CoPilot for FQN inference, but this study focuses on the large-scale evaluation, not human-CoPilot interaction.
Future work needs to explore the challenges and opportunities in human-CoPilot interaction discussed above.
%issues such as micro-level sensitivity and enhancement opportunities by various priori knowledge and human feedback.
Our experiments require crowd workers to manually request the CoPilot's completion.
To minimize human errors, we train crowd workers and ensure they can use CoPilot in the IDE successfully.
Furthermore, we automatically generate task inputs so crowd workers only need to perform minimum action (only need to press tab to request completion).
We also have automatic check to detect the missed task inputs.

% \subsubsection{Generalizability of Our In-context Learning Method}
Our prompt design and evaluation methodology is generic.
This study uses only Java code due to the dataset availability and the high effort to execute over 383K prompts.
Our follow-up work is to apply and evaluate the FQN prompts from this study on other libraries and programming languages such as Python, C++, and C\#.
As CoPilot is pre-trained with vast GitHub code in many programming languages, our follow-up work does not involve any gathering new code for model fine-tuning.

%% file: RelatedWork.tex
% \vspace{-2mm}
\section{Related Work}
% \vspace{-1mm}
%\zc{Some quick notes to organize related work in the following paragraphs:}

%PLMs and PLMs emergent abilities and data distribution properties; 
Many PLMs have been proposed in recent years, such as Bert~\cite{devlin2018bert}, T5~\cite{raffel2020exploring}, GPT3~\cite{Brown2020LanguageMA}, to name a few popular ones.
%\zq{These PLMs and their extensions have achieved the state-of-the-art in many NLP tasks, e.g., Closed Book Questions Answering~\cite{roberts2020much}, Translation~\cite{sennrich2015improving} and Reading Comprehension~\cite{lai2017race}.}
Researchers have investigated when and why PLMs achieve the superior performance in NLP tasks.
Wei et al.~\cite{Wan2022WhatDT} show that model scaling is crucial for emergent abilities of giant PLMs.
Chen et al.~\cite{Chan2022DataDP} identify three data distribution properties that drive emergent in-context learning capability in PLMs.
These findings inspire our adoption of in-context learning on CoPilot for FQN inference.

%Pretrained code model. The ways to transfer PLMs in the downstream tasks; FQN inference: huang ase2022 versus our in-context learning
PLMs have been extended to source code, following software naturalness~\cite{Devanbu2012OnTN, Allamanis2018ASO}, which produces pre-trained code models (PCMs), such as CodeBERT~\cite{Feng2020CodeBERTAP}, CodeT5~\cite{wang2021codet5}, Codex~\cite{chen2021evaluating} and CoPilot~\cite{GithubCopilot}.
These PCMs have significantly improve many SE tasks, such as code summarization~\cite{sun2022extractive}, fault prediction~\cite{rathore2019study}, vulnerability detection~\cite{li2021sysevr}, code translation~\cite{lachaux2020unsupervised}, FQN inference~\cite{Huang2022PrompttunedCL}.
% \zc{code summarization~\cite{sun2022extractive}, fault prediction~\cite{rathore2019study}, vulnerability detection~\cite{li2021sysevr}, code translation~\cite{lachaux2020unsupervised}, FQN inference~\cite{Huang2022PrompttunedCL} ... ??try to add some references for each task}.
There are two ways to transfer PLMs in the downstream tasks, supervised fine-tuning versus in-context learning (see Section~\ref{sec:backgroundknowledge}).
%Supervised fine-tuning initializes a task-specific model with the PLM weights, which is then further fine-tuned with downstream datasets.
%In contrast, in-context learning keeps the PLM frozen (i.e., no gradient update), but conditions the model on some task demonstrations to adapt its behavior in the downstream tasks.
All existing work on using PCMs adopts supervised fine-tuning, while our work is the first to adopt in-context learning in the SE task.

%In this setting, the pre-trained model weights are used to initialize a task-specific model which is then further fine-tuned.
%task-specific fine-tuning of the PLMs on the task-specific occurs mainly in the downstream heterogeneous.

%PLMs as neural knowledge base; probing pre-trained code model (zhang hongyu icse2021 (structural knowledge), ibm paper AST, asleep at the keyboard security); ours is SE factual knowledge; 
Many NLP studies~\cite{Roberts2020HowMK,petroni2019language, Jiang2020HowCW, heinzerling2020language} show PLMs can serve as neural knowledge bases, as opposed to symbolic knowledge bases.
SE researchers also attempt to probe different knowledge captured in PCMs, for example, structural knowledge like AST~\cite{zhang2019novel, AST}, semantic knowledge like code weakness~\cite{pearce2022asleep, li2018your}.
Some recent works investigate CoPilot's capability for code generation~\cite{vaithilingam2022expectation,pearce2022asleep} and code translation~\cite{Weisz2021PerfectionNR}, but they use only a small number of coding tasks.
Our study is the first to investigate the SE factual knowledge in PCMs at large scale.
In addition to FQNs, there are other types of SE factual knowledge often embedded in code, such as user credentials, text resources.
%Other SE factual knowledge (credentials, text resources), may suffer from unnecessary memorization and security leak (reference: extract training data from large language model; secret sharer)
%Although capturing FQNs is beneficial, 
Memorizing these facts may lead to serious security issues~\cite{carlini2021extracting, conrad2006secret}, as attackers may probe user credentials or commercial secrets in PLMs.
Although this is a real concern in adopting the models like CoPilot~\cite{pearce2021empirical}, no systematic study has been done to probe security-related factual knowledge in CoPilot.
Our research methodology can be extended to this aspect.
% Copilot~\cite{??seeif-anyworkorblogslikethesetwo-forcodemodel}

%Human-PLM interactions: zhang tianyi chi2021 copilot usability; perfect not required, CoAuthor, WordCraft. Our future work: teach Copilot to fix FQN generation error.
%AI Chain: inspiring idea - adapt a PLM in different roles and assemble roles, rather than one model for all. our is a FQN adapter, envision what could be other SE adapters and how to combine them (this may be part of discussion section)
Although PLMs demonstrate strong language capabilities, they can never be perfect due to their probabilistic nature~\cite{vaithilingam2022expectation, pearce2022asleep, Weisz2021PerfectionNR}.
Vaithilingam et al.~\cite{vaithilingam2022expectation} evaluate the usability of CoPilot in 3 simple programming tasks.
Our interaction experience with CoPilot echoes their findings, but we conduct large-scale experiments on a specific type of SE factual knowledge.
%study identifies various micro-level sensitivities of Copilot and human-fixable errors in FQN inference.
%It is a promising direction to investigate how to incorporate machine intelligence of PLMs and human intelligence.
%incorporation of machine intelligence of PLMs and human intelligence.
The imperfection of PLMs calls for innovative human-AI collaboration.
The tools like CoAuthor~\cite{lee2022coauthor} and WordCraft~\cite{coenen2021wordcraft} support human-GPT collaborative writing.
The quality of such collaborative writing is subjective.
However, human-CoPilot collaboration in SE tasks would demand objective outcomes (e.g., FQN correctness in our study), which demand innovative interaction designs for human-AI co-learning~\cite{Weisz2021PerfectionNR,lee2022coauthor}.
% co-learning~\cite{perfectnotrequired,googe-human-centric-ML-blog}

A promising paradigm for human-AI interaction is PLM recursion~\cite{wu2022ai} and AI chain~\cite{vaithilingam2022expectation}.
This paradigm adopts a divide-and-conquer strategy, in which the same PLM can be adapted to play different roles, and these roles and human roles can be linked to perform complex tasks.
Our FQN inference on CoPilot plays a specific role in code reuse.
We encourage the community to design other roles on the shoulder of PCMs and incorporate these roles to support complex coding tasks.

%% file: Conclusion.tex
\section{Conclusion}

This paper presents a lightweight in-context learning method on CoPilot for FQN inference in partial code, and designs a research methodology to evaluate the extent and characteristics of FQN knowledge in CoPilot and identify effective prompt designs and conditions to retrieve the FQN knowledge.
Our experiments confirm that CoPilot stores a large amount of priori knowledge of FQNs which can be accurately retrieved through zero or a few examples of task demonstrations.
Furthermore, in-context learning on CoPilot for FQN inference has no technical barrier (except for the CoPilot account cost) to deploy, as it does not require any code parsing or model tuning.
Our work demonstrates the benefits and practicality of standing on the shoulder of frozen giant PCMs and using the SE factual knowledge these models store in the SE tasks.
In the future, we will extend our FQN inference approach to more programming languages, and extend our research methodology to other SE factual knowledge (e.g., privacy and proprietary information in code).
We will investigate novel human-CoPilot interaction design to enhance human-AI collaboration in the SE tasks.

\section{Acknowledgements}
The work is partly supported by the National Nature Science Foundation of China under Grant (Nos. 62262031, 61902162, 61862033), the Nature Science Foundation of Jiangxi Province (20202BAB202015), the Graduate Innovative Special Fund Projects of Jiangxi Province (YC2021-S308, YC2022-S258).